\documentclass{aa}
\usepackage[usenames,dvipsnames]{xcolor}

\usepackage{graphicx}
\usepackage{amsmath, bm}
\usepackage{natbib}
\bibpunct{(}{)}{;}{a}{}{,} 

\usepackage{tikz}
\usepackage[percent]{overpic}
\usepackage{txfonts}
\usepackage{longtable}
\usepackage{listings}
\usepackage{color}
\usepackage{textcomp}
\usepackage{multirow}

\defcitealias{Cantat-Gaudin2020b}{CGa20}
\defcitealias{Tarricq2022}{T22}

\def\Gaia{{\it Gaia }}

\usepackage{soul}
\usepackage[normalem]{ulem}  

\begin{document}

\title{The multiplicity fraction in 202 open clusters from {\it Gaia}\thanks{The results are only available in electronic form
at the CDS via anonymous ftp to cdsarc.cds.unistra.fr (130.79.128.5)
or via https://cdsarc.cds.unistra.fr/cgi-bin/qcat?J/A+A/.}}

\author{J. Donada\inst{1,2,3}, 
        F. Anders\inst{1,2,3}, 
        C. Jordi\inst{1,2,3}, 
        E. Masana\inst{1,2,3}, 
        M. Gieles\inst{1,2,3,4}, \\ 
        G. I. Perren\inst{5}, 
        L. Balaguer-Núñez\inst{1,2,3}, 
        A. Castro-Ginard\inst{6},
        T. Cantat-Gaudin\inst{7},
        L. Casamiquela\inst{1,2,3,8}
        }
    
\institute{Departament de Física Quàntica i Astrofísica (FQA), Universitat de Barcelona (UB),  Martí i Franquès, 1, 08028 Barcelona, Spain
    \email{jdonada@fqa.ub.edu}
    \and{Institut de Ciències del Cosmos (ICCUB), Universitat de Barcelona (UB), Martí i Franquès, 1, 08028 Barcelona, Spain}
    \and{Institut d'Estudis Espacials de Catalunya (IEEC), Gran Capità, 2-4, 08034 Barcelona, Spain}
    \and{Catalan Institution for Research and Advanced Studies (ICREA), Passeig Lluís Companys 23, E-08010 Barcelona, Spain}
    \and{Instituto de Astrofísica de La Plata (IALP-CONICET), 1900 La Plata, Argentina}
    \and{Leiden Observatory, Leiden University, Niels Bohrweg 2, 2333 CA Leiden, The Netherlands}
    \and{Max-Planck-Institut für Astronomie, Königstuhl 17, D-69117 Heidelberg, Germany}
    \and{GEPI, Observatoire de Paris, PSL Research University, CNRS, Sorbonne Paris Cité, 5 place Jules Janssen, 92190 Meudon, France}
           }

\date{Received \today; accepted ...}

  \abstract{In this study, we estimate the fraction of binaries with high mass ratios for 202 open clusters in the extended solar neighbourhood (closer than 1.5 kpc from the Sun). This is one of the largest homogeneous catalogues of multiplicity fractions in open clusters to date, including the unresolved and total (close-binary) multiplicity fractions of main-sequence systems with mass ratio larger than $0.6_{ -0.15}^{+0.05}$. The unresolved multiplicity fractions are estimated applying a flexible mixture model to the observed \Gaia colour-magnitude diagrams of the open clusters. Then we use custom \Gaia simulations to account for the resolved systems and derive the total multiplicity fractions. The studied open clusters have ages between 6.6 Myr and 3.0 Gyr and total high-mass-ratio multiplicity fractions between 6\% and 80\%, with a median of 18\%. The multiplicity fractions increase with the mass of the primary star, as expected. The average multiplicity fraction per cluster displays an overall decreasing trend with the open cluster age up to ages about 100 Myr, above which the trend increases. Our simulations show that most of this trend is caused by complex selection effects (introduced by the mass dependence of the multiplicity fraction and the magnitude limit of our sample). Furthermore, the multiplicity fraction is not significantly correlated with the clusters' position in the Galaxy. The spread in multiplicity fraction decreases significantly with the number of cluster members (used as a proxy for cluster mass). We also find that the multiplicity fraction decreases with metallicity, in line with recent studies using field stars.
  }

\titlerunning{The multiplicity fraction in open clusters}
\authorrunning{J. Donada et al.}
\maketitle

\section{Introduction}
\label{sec:introduction}

The star formation mechanism leads to the creation of a significant fraction of pairs of stars and higher-order hierarchies \citep[e.g.][]{Tokovinin2020}. Determining how frequent those multiple stellar systems are is not straightforward because of the observational difficulties in identifying close pairs and distant binary systems  \citep[e.g.][]{Abt1976}.

Open star clusters (OCs) are highly suitable for studying stellar binary systems, because OC member stars have, to first order, the same distance, age, initial chemical composition, and foreground extinction. Although most of their binaries are unresolvable in images, they can be identified in the colour-magnitude diagram (CMD): they move above the main sequence (MS) depending on their mass ratio $q$\footnote{$q=M_{2}/M_{1}$, where $M_1$ is the mass of the primary component (i.e., the one with the larger initial mass), $M_2$ is the mass of the secondary and $0\leq q \leq 1$.}, with the equal-mass binaries being $0.753$ mag brighter than their MS counterparts. The binary fraction in open clusters has therefore been an active field of study for decades \citep[e.g.][]{Mermilliod1989, Sandhu2003, Bica2005, Sharma2008}.

The characterisation of the multiplicity fraction in OCs and field stars is relevant to several branches of astrophysics \citep[for reviews see e.g.][]{Duchene2013, Moe2017, Offner2022}. Correctly accounting for unresolved binary stars is essential in determining the total mass and stellar mass function of OCs \citep[e.g.][]{Kroupa2001,Borodina2019, Rastello2020}. They also influence the OC’s dynamical evolution, providing relevant information about the outcome of star forming processes in different environments and
constraints on the initial OC’s state \citep[e.g.][]{Li2020}. Furthermore, binaries are responsible for high rates of stellar collisions \citep{2004MNRAS.352....1F, Gonzalez2021}, black-hole mergers \citep{2002ApJ...572..407B} and tracers of intermediate-mass black holes \citep{Aros2021}, thus proving relevant for high-energy astrophysics and gravitational-wave studies \citep{Banerjee2022}. 

Since typically Galactic OCs are disrupted on a time scale of a few hundred Myr \citep{2003ARA&A..41...57L, Lamers2005}, they tend to be relatively young \citep{1971Ap&SS..13..300W, Anders2021}. For young massive clusters in the local universe, the measured multiplicity fractions are close to those of field stars \citep{Gonzalez2021}. 
In this paper we attempt to measure the high mass-ratio ($q>0.6$) multiplicity fraction for an unprecedented number of Galactic OCs located closer than 1.5 kpc to the Sun in an automated fashion, making use of the exquisite \Gaia photometry \citep{Gaia-fot-edr3}. 
Our main input datasets are the recent \Gaia OC catalogues published by \citet[][hereafter T22]{Tarricq2022} and \citet[][hereafter CGa20]{Cantat-Gaudin2020b}. 

The paper is structured as follows: In Sect. \ref{sec:problem} we introduce the problem of estimating the multiplicity fraction in OCs. 
The treatment of the \Gaia data is described in Sect. \ref{sec:data}. In Sect. \ref{sec:mcmc} we explain our method to measure the multiplicity fraction in each cluster. In the following Sect. \ref{sec:sims} we validate the method (and determine the effective mass-ratio limit, $q_{\rm lim}$, to which our method is sensitive) using custom OC simulations performed with the \emph{Gaia Object Generator} \citep{Luri2014}. 
We present and discuss our results in Sect. \ref{sec:results} and conclude in Sect. \ref{sec:conclusions}.

\section{Setting the problem}\label{sec:problem}

The multiplicity fraction $f_{b}$ of a stellar population (or, obviating higher-order systems, the binary fraction), is defined as:
\begin{equation}
    f_{b}=\frac{B+T+...}{S+B+T+...},
\end{equation}
where $S$ is the number of single stars, $B$ the number of binary systems, $T$ the number of triple systems, and so on. The principal aim of this study is to estimate the $f_{b}$ of MS unresolved multiple systems in OCs, using {\it Gaia}’s $G$ vs. $(BP-RP)$ CMD. 

An unresolved binary system composed of two identical main-sequence stars ($q=1$) has the same colour but twice the luminosity of an equivalent single star, and appears in the OC’s CMD $0.753$ mag brighter than the equivalent single star location (irrespective of the photometric band). 
A system with two unequal main-sequence components ($q<1$) has a combined colour that is redder than the colour of the primary component, and a combined luminosity greater than the one of a single star but less than the one corresponding to the equal-mass binary system. Therefore, such a system is displaced in the CMD both upwards (but not reaching 0.753 mag) and to the right relative to the main-sequence position of the primary component \citep[see e.g. Fig. 1 in][]{2012A&A...540A..16M}. 
With $q$ increasing from 0 to 1, the position of the binary system with respect to the main-sequence position of the primary component always decreases in magnitude, but for $0<q<q_{\rm crit}$ moves towards redder colours, and for $q_{\rm crit}<q<1$ towards the blue again \citep{Maeder1974, Hurley1998}. The value of $q_{\rm crit}$ depends on the mass of the primary component, caused by the fact that the shifts in magnitude and colours of a system of $q < 1$ depend on the magnitude and colour difference of the two components. Hence, the binary sequences of a certain $q < 1$ do not have a constant separation from the equal-mass binary sequence throughout all
the MS. In most of the CMD MS range, there is not much separation between the $q = 0$ and $q = 0.5$ isochrones, so the low-$q$ unresolved binaries create an overdensity near the $q = 0$ sequence (single-star main sequence, SS). On the other hand, the secondary
sequence (binary sequence, BS) observed in most CMDs above the SS comprises not
only equal-mass systems, but also high-$q$ unresolved binaries. Hence, the approximate location of a system on the binary sequence does not imply that it is an equal-mass binary (as first shown in \citealt{Hurley1998}).

$q_{\rm crit}$ does not exactly coincide with the smallest $q$ for which a binary system lies on the binary sequence ($q_{\rm lim}$). From the observational perspective, $q_{\rm lim}$ is a more interesting parameter to consider: it is the mass ratio above which a binary system can be detected as such from photometry alone. 
Other effects like differential reddening, stellar rotation, and blends can potentially also displace a single star from its location in the MS towards redder and/or brighter positions mimicking multiplicity. The effect of blended sources is negligible for the vast majority of {\it Gaia} OCs in the solar neighbourhood. Even in cases of extremely dense fields such as for the distant massive open clusters NGC 1805 and NGC 1818 located in the Large Magellanic Cloud, \citet{Li2013} found that the effect of blending is negligible for stars brighter than $V=18$ mag (see their Figs. 6 and A5).

\section{{\it Gaia} cluster member data} \label{sec:data}

\subsection{OC memberships and parameters}

Since we are interested in an OC census that is representative for the extended solar neighbourhood, we pre-select all 368 local ($d<1.5$ kpc) OCs from the {\it Gaia} EDR3-based \citep{Gaia-EDR3} catalogue of \citet{Tarricq2022}. For OCs younger than 50 Myr (for which \citetalias{Tarricq2022} do not provide improved membership lists), we supplement this dataset with the {\it Gaia} DR2-based catalogue \citep{Gaia-DR2} of \citetalias{Cantat-Gaudin2020b}, pre-selecting its 302 OCs younger than 50 Myr and closer than $1.5$ kpc. In both cases, we use the homogeneously derived astrophysical parameters (age, distance, extinction) published in \citetalias{Cantat-Gaudin2020b}. We have, therefore, a preliminary sample of 670 local OCs, where the membership lists are taken from \citet{Tarricq2022} for OCs older than 50 Myr and from \citetalias{Cantat-Gaudin2020b} for OCs younger than 50 Myr. 377 out of the 670 OCs have at least 30 MS members and a MS that extends over at least 1 mag in $BP-RP$ and are considered for a detailed analysis (see below).

\begin{figure}
\begin{center} 
\includegraphics[width=.495\textwidth]{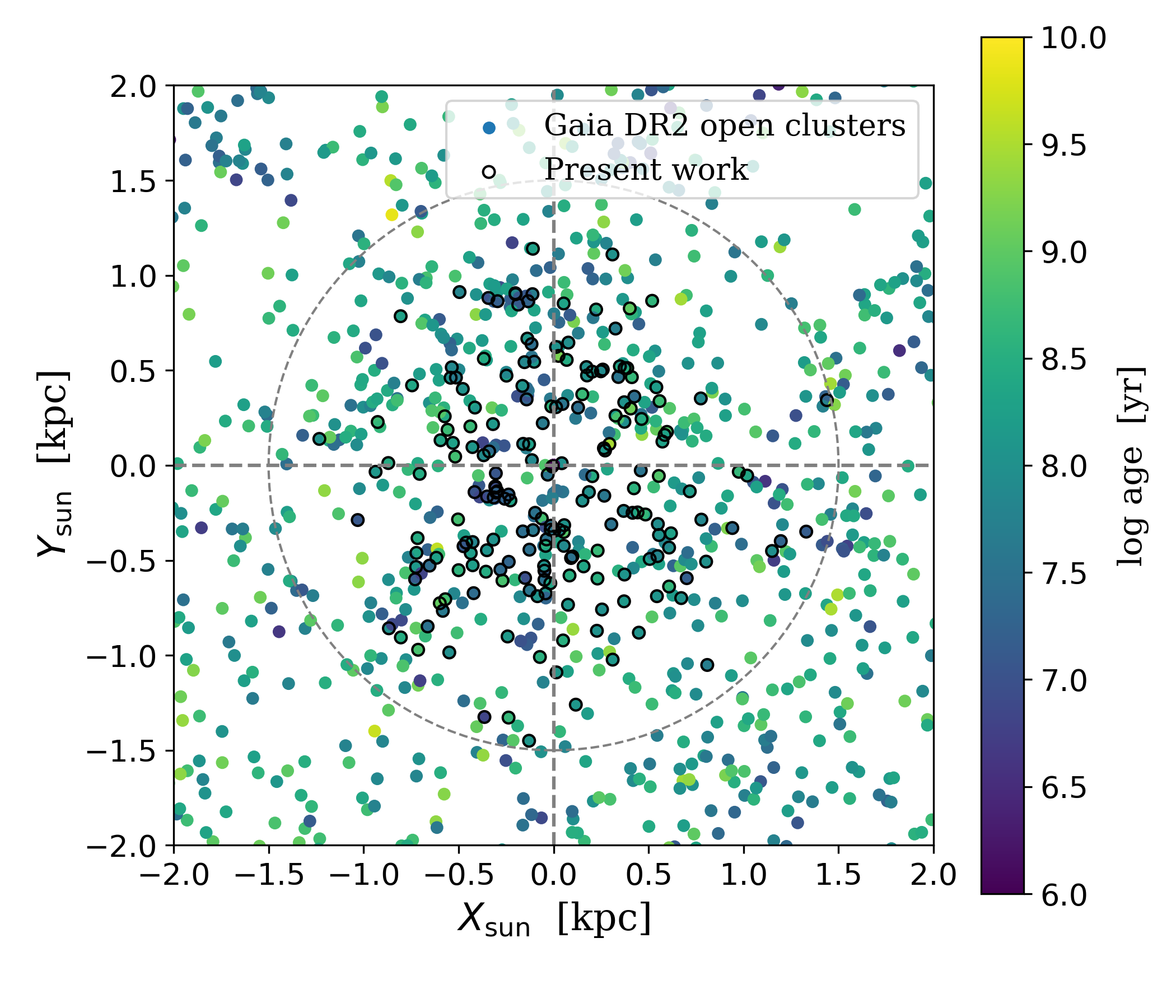}\\
\includegraphics[width=.495\textwidth]{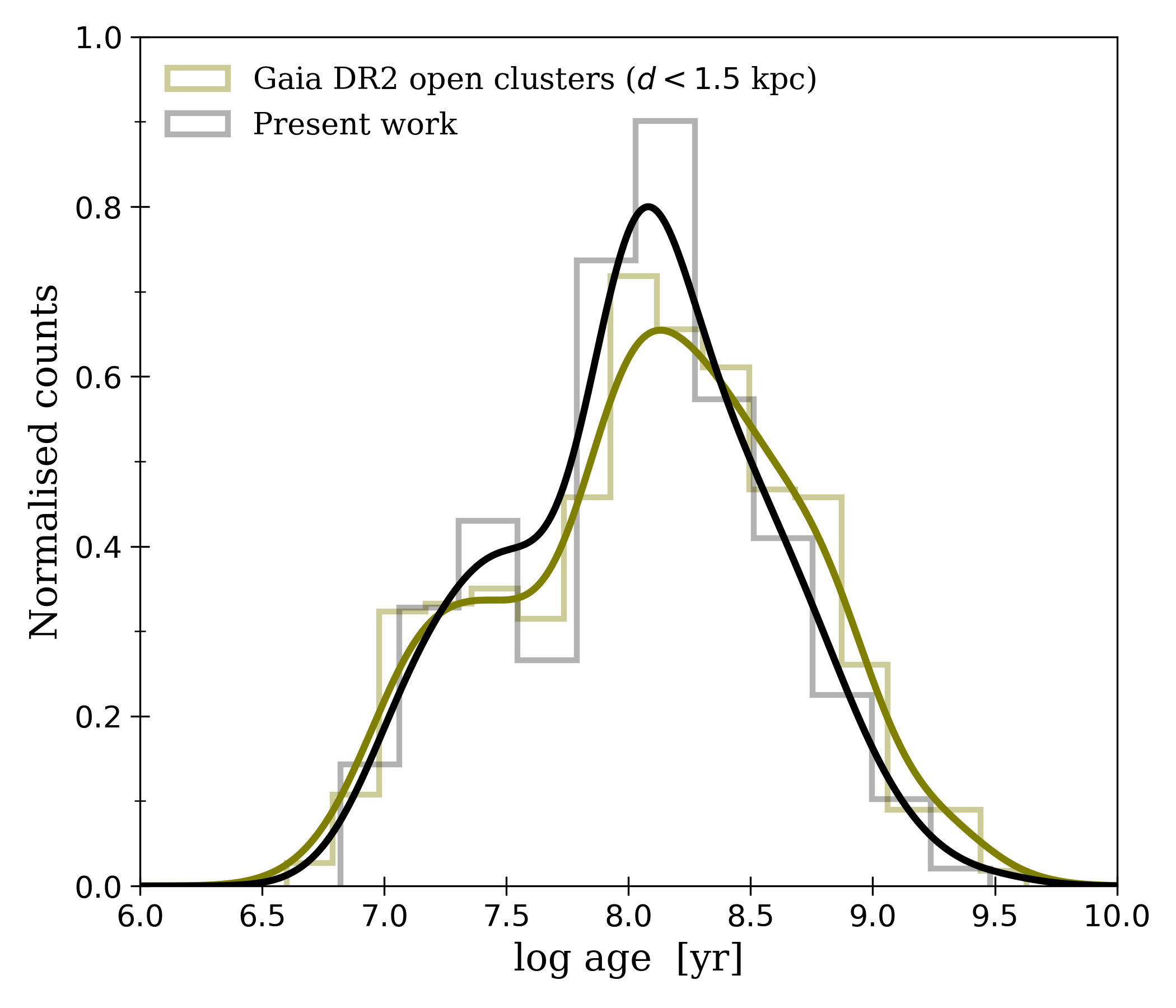}\\
\caption{\label{fig:xymap} 
\emph{Top panel}: spatial distribution of the open clusters in the {\it Gaia} DR2-based catalogue of \citetalias{Cantat-Gaudin2020b} (points colour-coded by age) and of the 202 OCs in our final sample (points encircled by a black circumference) in heliocentric Cartesian coordinates. The Galactic Centre is towards the right. \emph{Bottom panel}: Normalised age distributions for the local OCs in \citetalias{Cantat-Gaudin2020b} (738 OCs with $d<1.5$~kpc, olive histogram and kernel-density estimates; 202 of them constitute our final sample of OCs whose membership list is taken from either \citetalias{Tarricq2022} or \citetalias{Cantat-Gaudin2020b} depending on the age) and for our final sample of 202 OCs (black histogram and kernel-density estimates).}

\end{center}
\end{figure}

Both membership catalogues are based on astrometric data limited to $G<18$.\footnote{\cite{Penoyre2022} showed how the proper motions and parallax of unresolved binaries can be biased depending on the period. This can sometimes result in true member stars being misclassified as field stars.} Regarding the considered members of each OC, we have used all the members in the \citetalias{Cantat-Gaudin2020b} catalogue that have membership probabilities $\geq 70\%$ and have selected those of \citetalias{Tarricq2022} with membership probabilities above the same threshold value. However, as the criteria for the membership probability assignation are not equal, establishing this common threshold does not make the selection exactly equivalent for both catalogues. 
After the further selections described in Sect. \ref{sec:mcmc}, a final sample of 202 OCs is obtained. Figure \ref{fig:xymap} shows their spatial and age distribution.%

\begin{figure}
\begin{center} 
\includegraphics[width=.495\textwidth]{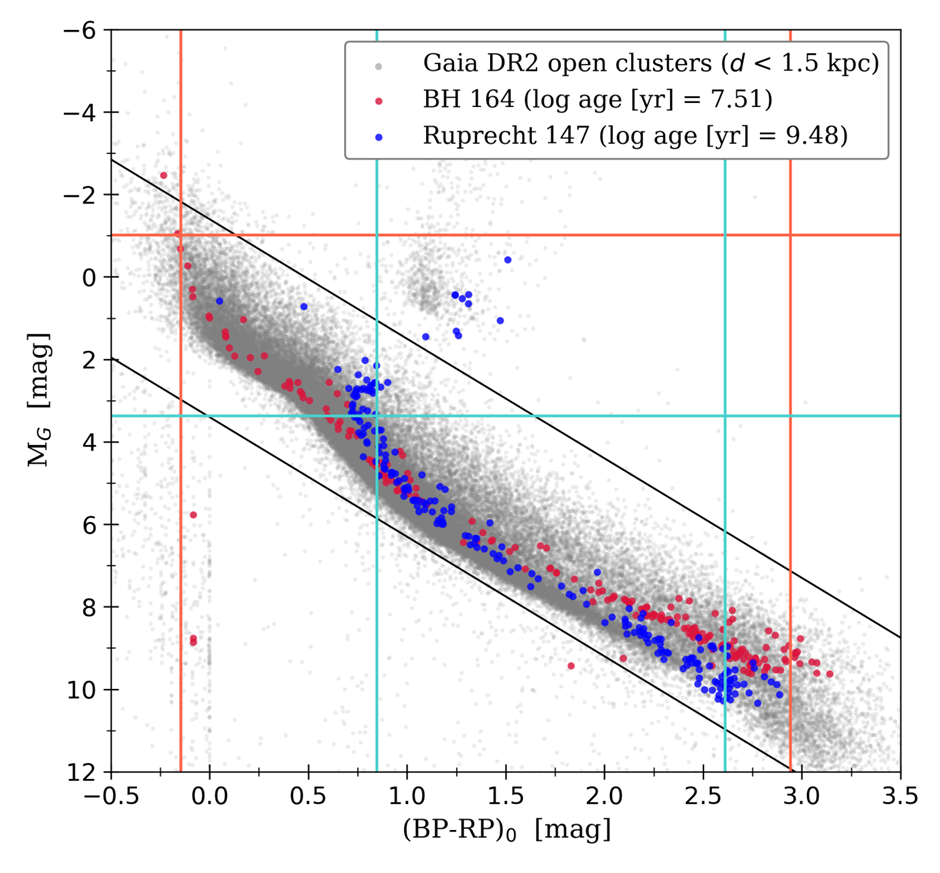}\\
\caption{\label{fig:cmd_cuts} 
Two examples showing the selection of main-sequence members from the dereddened CMD. The black lines indicate the exclusion mainly of giants and white dwarfs. The red and blue lines indicate the cuts to exclude MS turn-off stars and to avoid incompleteness at the red end for the BH 164 and Ruprecht 147 (respectively).
} 
\end{center}
\end{figure}

\subsection{Selection of OC main-sequence members}\label{sec:ms_select}

We adopted a homogeneous selection of the main sequence of each OC in our sample. 
First of all, following \citetalias{Cantat-Gaudin2020b} the absolute magnitude $M_G$ and intrinsic colour index $(BP-RP)_0$ were calculated for each cluster member (excluding the ones with missing $BP$ and/or $RP$ photometry):
\begin{align}
\label{eqn:M_G}
    M_G   &= G - \mu -0.89\cdot A_V, \\
\label{eqn:BP_RP_intri}
(BP-RP)_0 &= (BP - RP) - \frac{0.89}{1.85}\cdot A_V,
\end{align}
where $G$ is the {\it Gaia} $G$-band mean magnitude, $\mu$ is the distance modulus from \citetalias{Cantat-Gaudin2020b}, and $A_V$ is the OC's visual extinction (also from \citetalias{Cantat-Gaudin2020b} and corrected by a factor in equation \ref{eqn:M_G} to obtain the extinction in the $G$-band, and by a different factor in equation (\ref{eqn:BP_RP_intri}) to obtain the $E_{(BP-RP)}$ colour excess). 
With these two quantities, we can now display the dereddened CMD for the OCs in our parent sample. We now derive restrictions (simple cuts in the dereddened CMD) for each member of a particular OC to be considered a MS member. The procedure (described below) is illustrated for two OCs of different ages in Fig. \ref{fig:cmd_cuts}.

As our multiplicity fraction determination involves only MS
stars, two straight parallel lines which lie above and below the MSs of all clusters can be drawn in the CMD. This results in a first rough selection that excludes red-clump stars, white
dwarfs, and some extreme outliers (black lines in Fig. \ref{fig:cmd_cuts}):
\begin{equation}
    2.9\cdot(BP-RP)_0 -1.4 < M_G < 2.9\cdot (BP-RP)_0+3.4
\end{equation}

\begin{figure*}
\centering
    \begin{tikzpicture}
       \node[anchor=south west,inner sep=0] (image) at (0,0) {\includegraphics[width=0.99\textwidth]{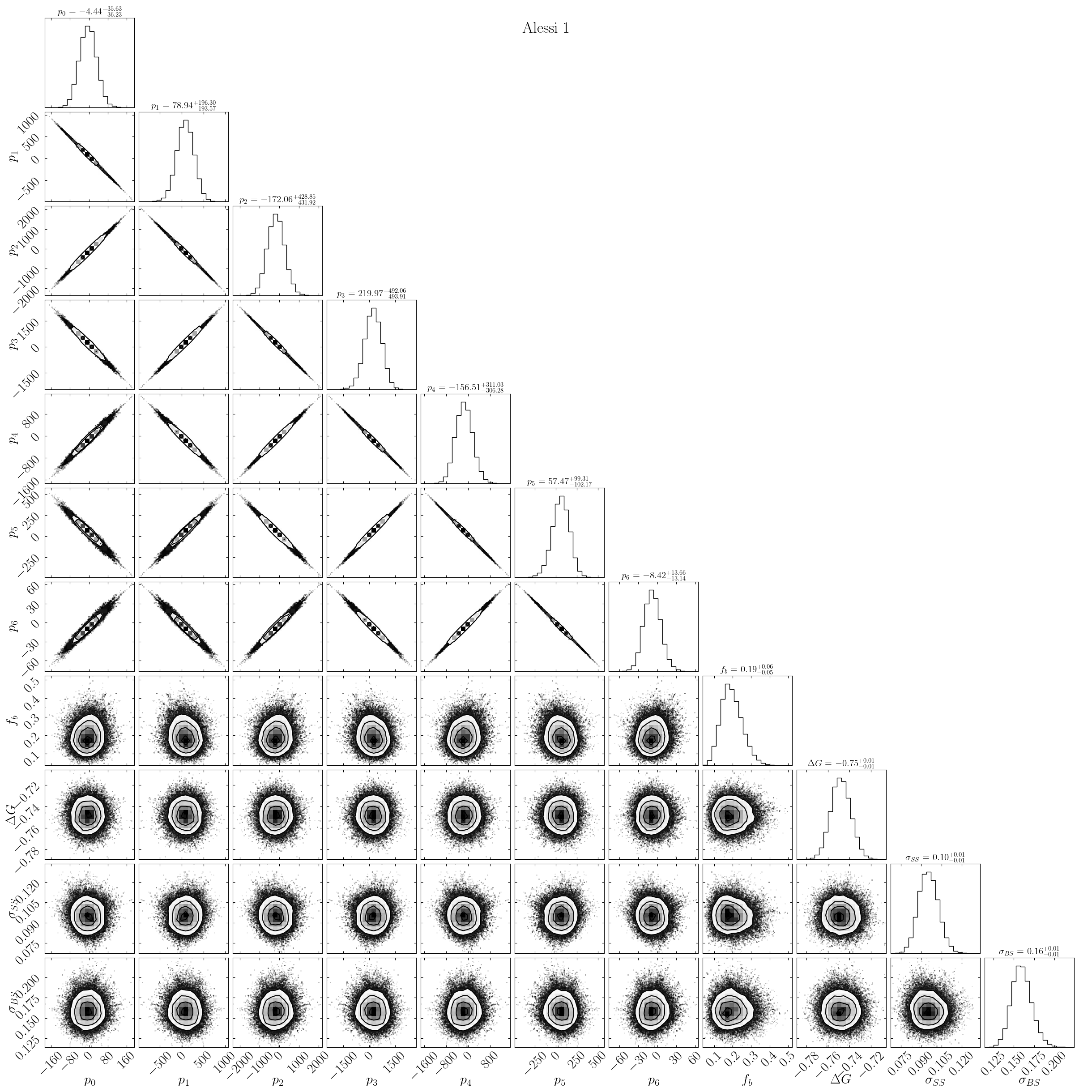}};
       \begin{scope}[x={(image.south east)},y={(image.north west)}]
       \node[anchor=south west,inner sep=0] (image) at (0.382,0.655) {\includegraphics[width=0.4\textwidth]{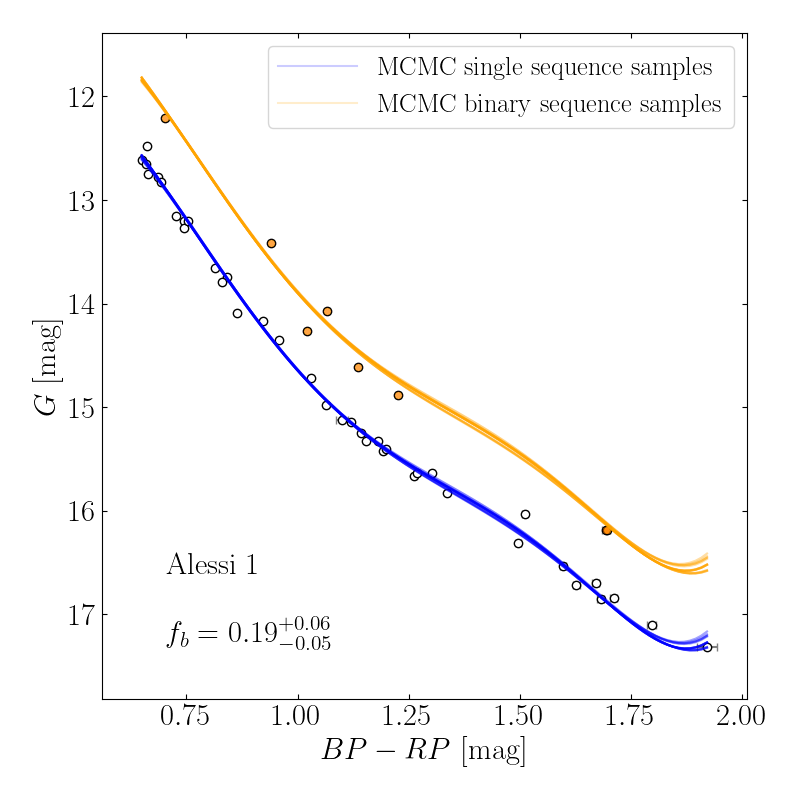}
        \includegraphics[width=0.2\textwidth]{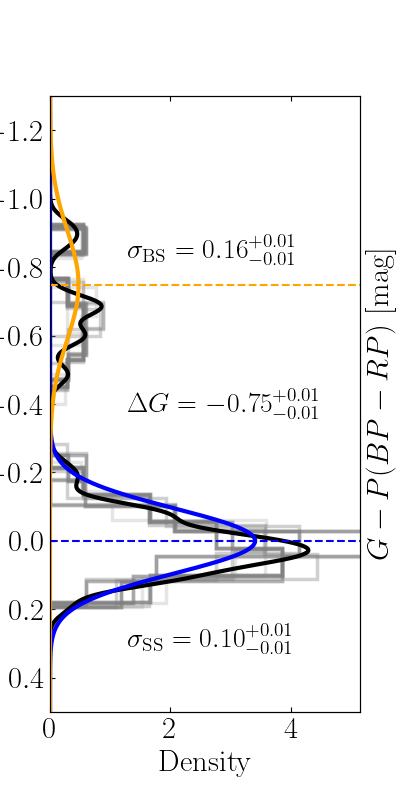}
       };
       \end{scope}
    \end{tikzpicture}
\caption{MCMC fit for Alessi 1, which has 46 selected MS members. \emph{Lower left panels}: {\tt corner} plot of the posterior. The first 7 parameters correspond to the polynomial coefficients $p_i$, while the last four describe the unresolved multiplicity fraction $f_b$, the separation in $G$ magnitude between the binary sequence and the single-star sequence $\Delta G$, the intrinsic $G$-magnitude spread of the single-star sequence $\sigma_{SS}$, and the one of the binary sequence $\sigma_{BS}$. The main outcome of the fit is the unresolved multiplicity fraction $f_b$, all the other parameters are considered nuisance parameters.
\emph{Top right panels}: CMD overplotted by 30 random samples of the posterior SS and BS, where the members $>3\sigma_{SS}$ brighter than the fitted SS polynomial are coloured in orange (left). The corresponding residuals with respect to the polynomial fit are also shown as histograms and kernel-density estimates, together with the resulting Gaussian mixture fit (right).}
\label{fig:mcmc_fit_corner_alessi1} 
\end{figure*}

Another restriction is required to exclude members which are already evolving towards giant stars. In order to estimate the intrinsic colour index of the MS turn-off (MSTO), we fitted an interpolation polynomial $f(\log \tau)$ to the $(BP-RP)_0$ dependence on the age ($\tau$) for the bluest point of the MSs of 15 different PARSEC 1.2S isochrones \citep{Bressan2012} of solar metallicity and with ages in years in the range $\log \tau\in[6.6, 9.8]$. So, for each OC, the dereddened colour of the MSTO was obtained as a function of its age as $(BP-RP)_0^{TO}=f(\log \tau)$.

Then, MSTO and blue-straggler stars were excluded by only considering members at least 0.2 mag redder than the MSTO (i.e., members with $M<M_{TO}$ were selected):
\begin{equation}
    (BP-RP)_0 > (BP-RP)_0^{TO} + 0.2 
\end{equation}
and also imposing the extra condition that their magnitude was fainter than the one of the bluest member selected in the previous step minus 0.75 mag:
\begin{equation}
    M_G \geq M_G((BP-RP)_{0,{\rm bluest\ member}})-0.75
\end{equation}
Such conditions were found to succeed in excluding from the study not only the MSTO members, but also those extreme MS members for which the binary sequence approaches the single-star main sequence and eventually intersects it. For these cases our mixture-model algorithm (see Sect.~\ref{sec:mcmc}) would not work because it is based on assuming a roughly constant separation between both sequences.

Finally, the redder MS members of each OC were excluded to avoid an overestimation of the multiplicity fraction due to the magnitude limit ($G<18$) in the parent cluster membership catalogues. We therefore selected only stars 0.2 mag bluer than the reddest star:
\begin{equation}
    (BP-RP)_0 \leq (BP-RP)_{0, {\rm reddest\ member}} - 0.2. 
\end{equation}

The subsequent study is carried out only for those OCs having a number of retained MS members larger or equal than 30, and, at the same time, a MS length in  $(BP-RP)_0$ longer than or equal to 1 mag (computed as the rest of the maximum and minimum values of  $(BP-RP)_0$ within the selected members). 
We now have an input sample of 377 OCs closer than 1.5 kpc: 250 OCs older than 50 Myr with EDR3 memberships from \citetalias{Tarricq2022} and 127 OCs younger than 50 Myr with DR2 memberships from \citetalias{Cantat-Gaudin2020b}.

\section{Fitting the single-star and binary sequences}\label{sec:mcmc}

Studying stellar multiplicity in a statistically robust manner is a highly non-trivial task (e.g. \citealt{Duquennoy1991, Belloni2017}). The mass ratio of binaries, for instance, is interconnected with the primary mass, the orbital period, the eccentricity, and the system metallicity \citep{Offner2022}. 

Following the path laid out in \citet{Hogg2010}, we can write down a mixture model for the distribution of cluster members in the CMD. If the functional form of the single-star main sequence (SS) in the \Gaia CMD is perfectly known (i.e. we know the functional form $G_{SS}= f(BP-RP)_{SS}$ + scatter), the cluster CMD (at least the part sufficiently redder than the MSTO) can be described as a simple mixture model of two populations that are described by the same functional form (separated by a constant offset $\Delta G = G_{BS}-G_{SS}\simeq-0.75$ mag) plus their respective intrinsic scatter in $G$ magnitude. One population mostly accounts for single stars, resolved binaries and low-$q$ unresolved binaries, seen in the CMD as a widened single-star sequence approximately bottom bounded by the single-star sequence locus. The other population mostly accounts for unresolved binaries of $q_{\rm lim}<q \leq 1$, seen in the CMD as a widened binary sequence, BS, approximately bound by the equal-mass binary sequence locus.
In a Gaussian approximation for their scatter in $G$ magnitude, the unresolved $f_{b}$ is then just the weight of the binary sequence Gaussian: $f_{b}(q> q_{\rm lim})=\frac{w_{BS}}{w_{SS}+w_{BS}}=w_{BS}$ (if both weights are normalised: $w_{SS}+w_{BS}=1$). Assuming these distributions to be Gaussians is a first-order approach. 

\begin{figure*}
\includegraphics[width=.33\textwidth]{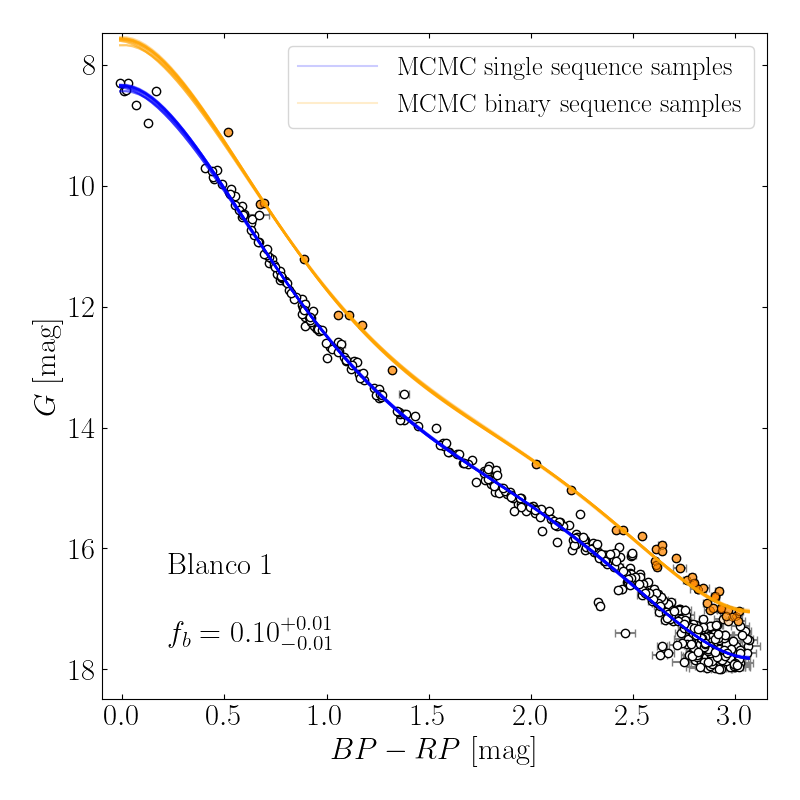} 
\includegraphics[width=.16\textwidth]{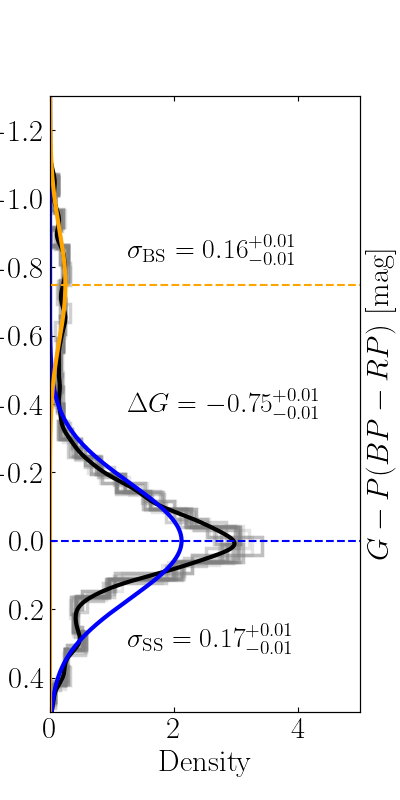} 
\includegraphics[width=.33\textwidth]{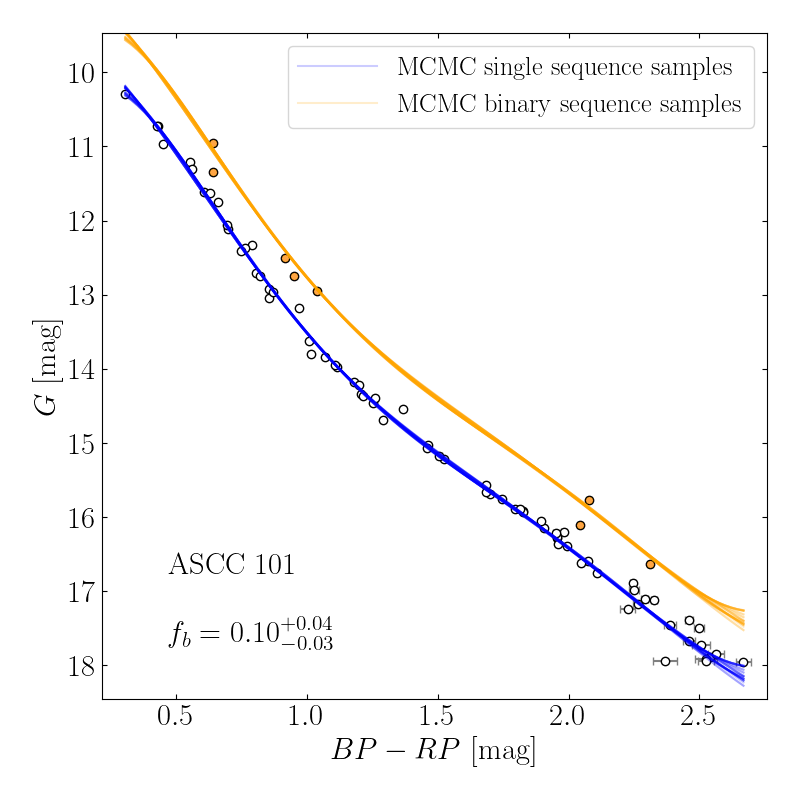} 
\includegraphics[width=.16\textwidth]{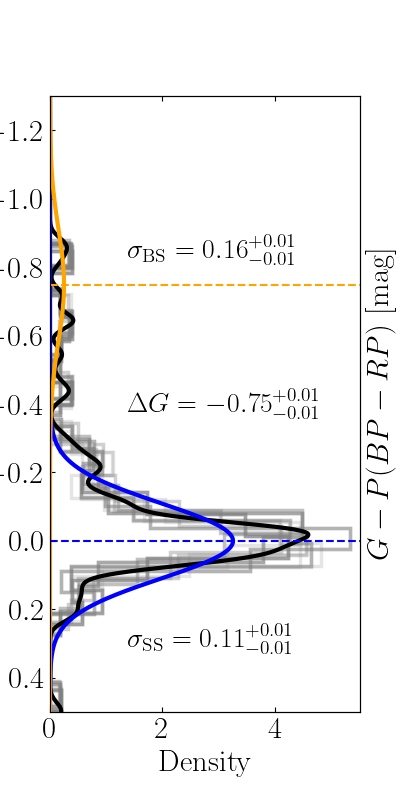} 
\includegraphics[width=.33\textwidth]{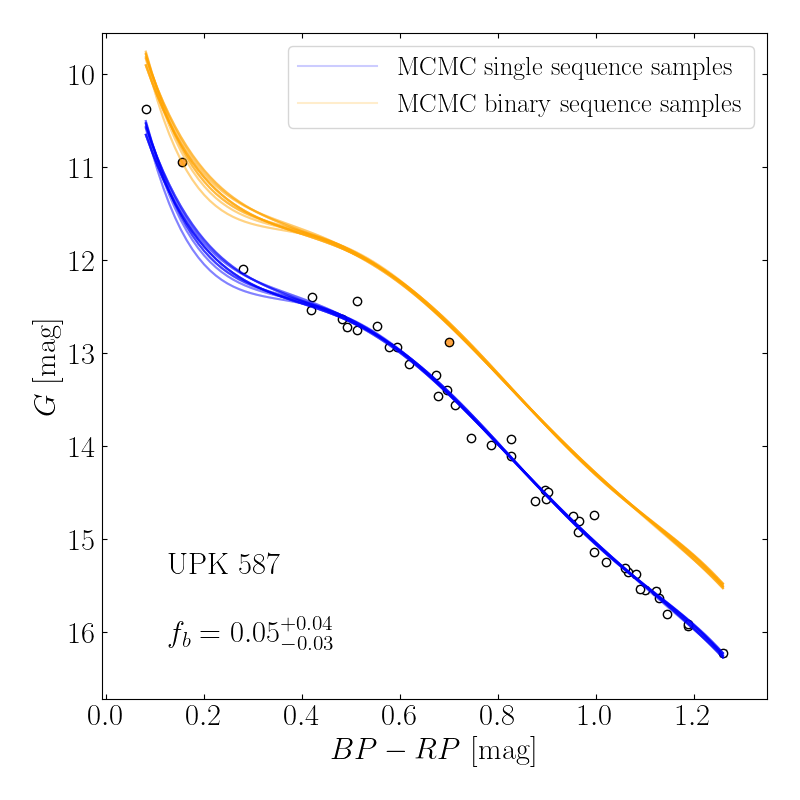} 
\includegraphics[width=.16\textwidth]{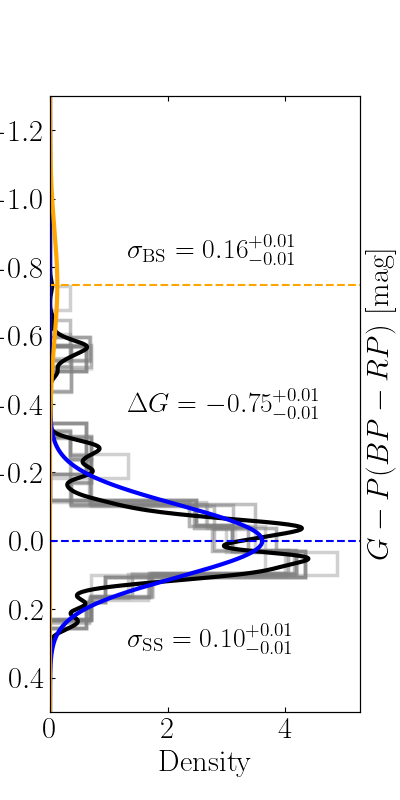} 
\includegraphics[width=.33\textwidth]{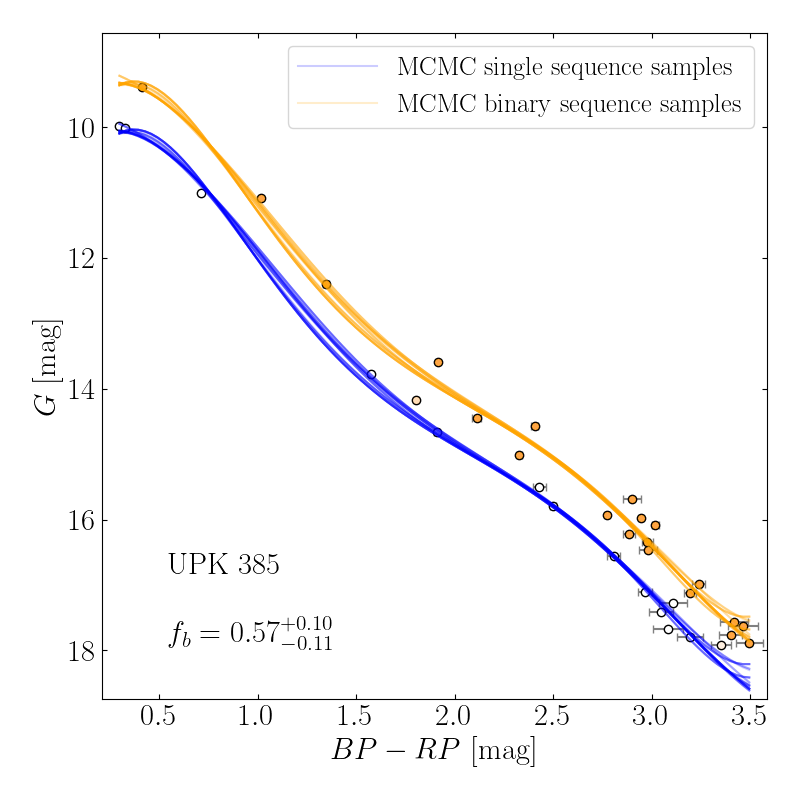} 
\includegraphics[width=.16\textwidth]{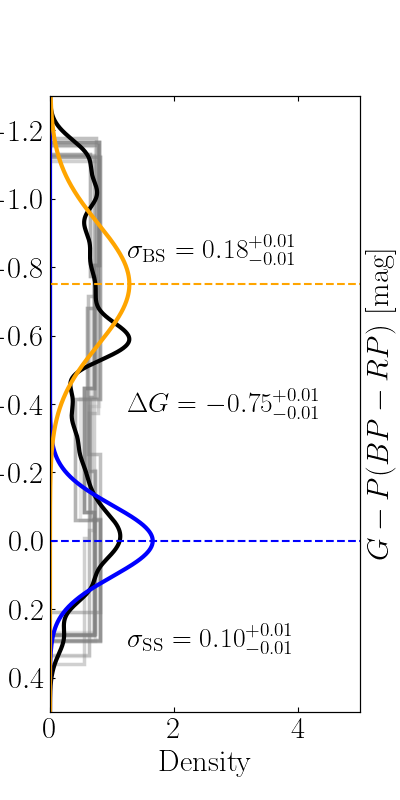} 
\includegraphics[width=.33\textwidth]{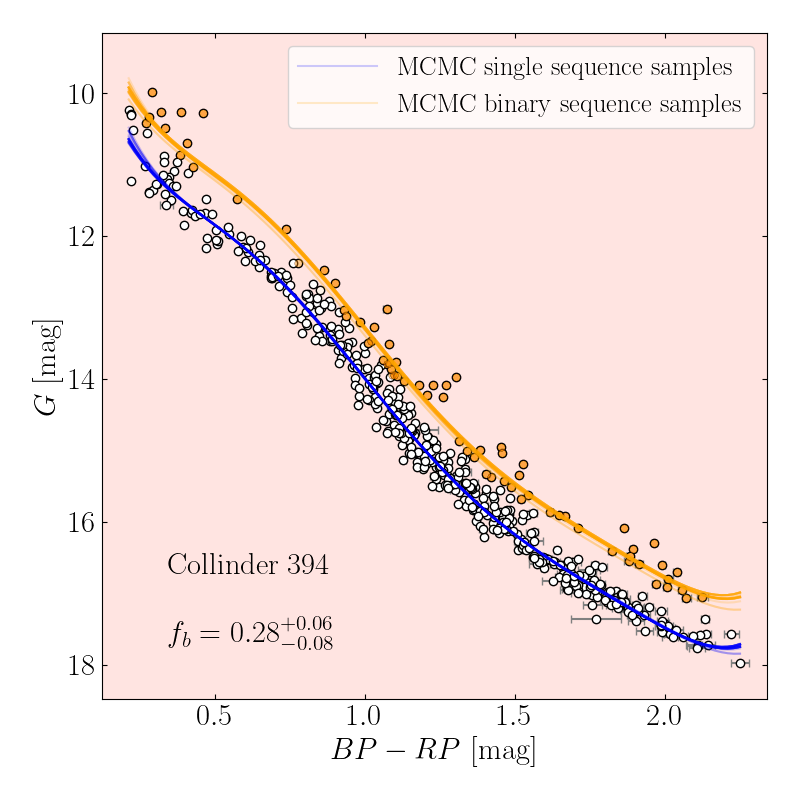} 
\includegraphics[width=.16\textwidth]{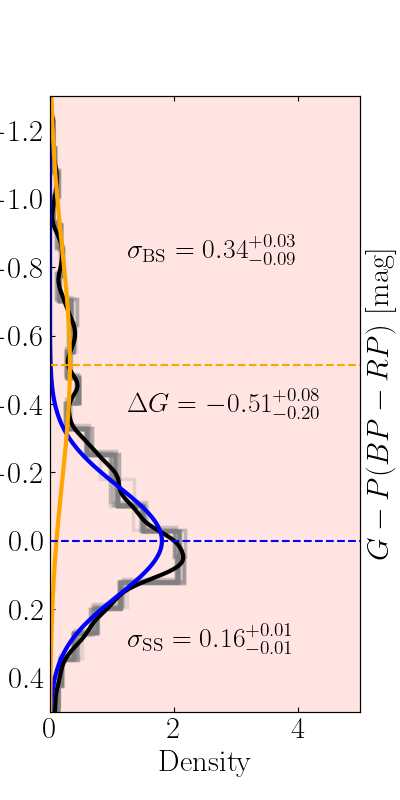} 
\includegraphics[width=.33\textwidth]{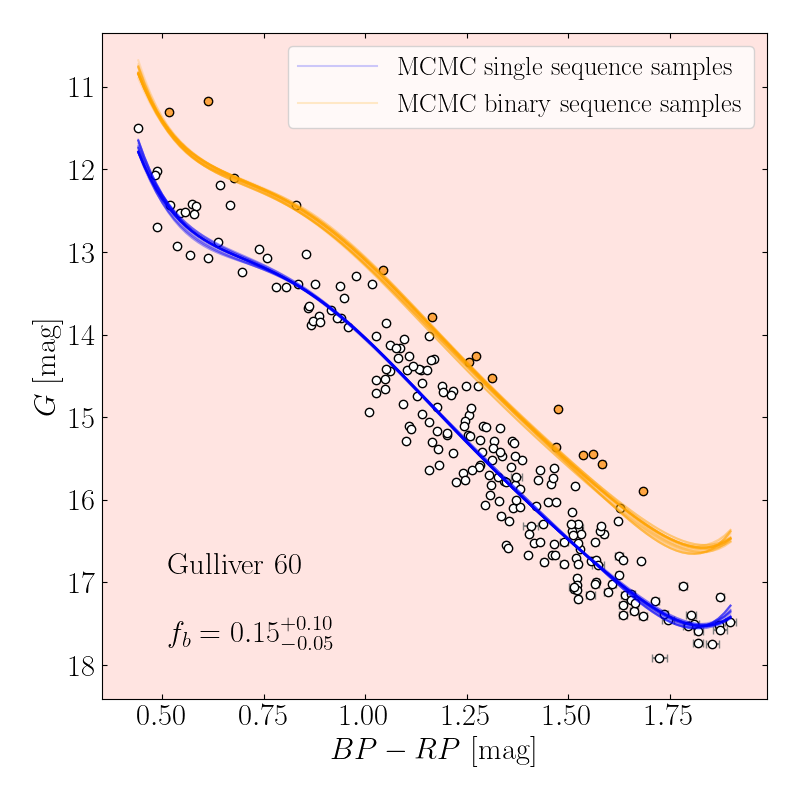} 
\includegraphics[width=.16\textwidth]{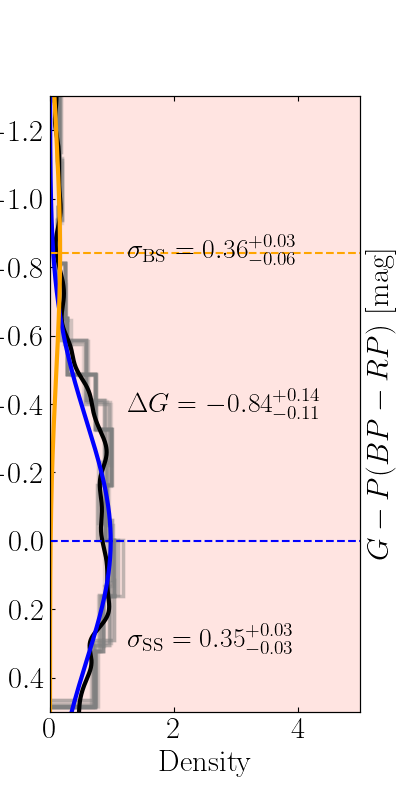} 
\caption{ 
Examples of the MCMC fits to the main sequences of several OCs (in the same style as Fig. \ref{fig:mcmc_fit_corner_alessi1} top right panels), illustrating the diversity of the binary sequences and the resulting diversity in the quality of the MCMC fits. The two examples marked by a light red background are excluded from the final catalogue because they do not fulfill the conditions described in Sect. 4.} 
\label{fig:mcmc_examples}
\end{figure*}

In fact, this is the integrated high-$q$ unresolved $f_b$ in the MSs of OCs, because two Gaussian distributions in $G$ magnitude are considered for all the MS systems. And, as each Gaussian has a fixed standard deviation, the $q$ range each of them comprises varies as a function of $M_1$; and hence the $q_{\rm lim}$ below which multiple systems are regarded as simple systems varies slightly as a function of $BP-RP$ \citep{Hurley1998}. We, however, provide a single mean $q_{\rm lim}$ value for each OC, having integrated over the colour index and the other dimensions introduced by $q$ dependencies. We also assume negligible contamination by field star interlopers, a valid assumption for the exquisite {\it Gaia} OC catalogues in the vast majority of cases; and triples and higher-order systems cannot be distinguished from binaries by our method (being included in the binary sequence population), so the estimated $f_{b}(q> q_{\rm lim})$ includes their contribution.

The likelihood for the CMD distribution below the turn-off can be described, therefore, by an arbitrary polynomial $P(BP-RP)$ with a small intrinsic $G$-magnitude scatter for the SS, $\sigma_{SS}$, and a parallel one (perhaps with a slightly larger intrinsic scatter $\sigma_{BS}$) for the BS. If we further assume the observational errors in $BP-RP$ colour and $G$ magnitude to be uncorrelated, the likelihood can be written as
\begin{align}
\label{eqn:likeli}
\begin{split}
 \ln \mathcal{L} = \sum_{i=1}^{N} \ln \Bigg[  &\frac{(1-f_b)}{\sqrt{2\pi(\sigma_{{\rm eff}, i}^2+\sigma_{SS}^2)}} \cdot {\rm e}^{-\frac{(M_{G_i} - P(BP-RP)_i)^2}{2(\sigma_{{\rm eff}, i}^2+\sigma_{SS}^2)}} \\ 
 &+ \frac{f_b}{\sqrt{2\pi(\sigma_{{\rm eff}, i}^2+\sigma_{BS}^2)}} \cdot {\rm e}^{-\frac{(M_{G_i} - P(BP-RP)_i + \Delta G)^2}{2(\sigma_{{\rm eff}, i}^2+\sigma_{BS}^2)}} \Bigg],
\end{split}
\end{align}
where, in analogy to Chapter 7 of \citet{Hogg2010}, we have introduced the effective uncertainties $\sigma_{{\rm eff}, i}$ (the combined uncertainty projected onto the polynomial fit), defined as
\begin{equation}
\begin{split}
    \sigma_{{\rm eff}, i}^2 &:= \frac{1}{1 + m_i} (\sigma_{G, i}^2 +  m_i\cdot \sigma_{(BP-RP)_i}^2) \\
    & {\rm with}\quad m_i := \left(\frac{\partial P(x)}{\partial x}\right)^2|_{{x=(BP-RP)}_i}.
\end{split}
\end{equation}

The likelihood in equation (\ref{eqn:likeli}) is thus a function of $p+5$ parameters, being $p$ the order of the fitted polynomial $P$ ($p+1$ nuisance parameters for the polynomial itself, the intrinsic width of the SS, $\sigma_{SS}$, the intrinsic width of the BS, $\sigma_{BS}$, the vertical offset between the two sequences $\Delta G$, and the cluster's high-$q$ unresolved multiplicity fraction $f_b$). 

Since we are interested in sampling the posterior PDF of this parameter space, we also need to impose some priors. To leave maximal possible freedom to the fitting algorithm, we do not impose priors on the polynomial coefficients and only quantify our a-priori knowledge of stellar evolution. In particular, we assume Gaussian priors for the vertical offset $\Delta G$ and the logarithm of the width of the SS and the BS, respectively:
\begin{align}
    p_1(\Delta G) &= \mathcal{N}(-0.75, 0.05)\\ 
    p_2(\log_{10} \sigma_{BS}) &= \mathcal{N}(-1, 0.2)\\ 
    p_3(\log_{10} \sigma_{SS}) &= \mathcal{N}(-0.8, 0.2)\\
    p_{\rm full}(\Delta G, \log_{10} \sigma_{SS}, \log_{10} \sigma_{BS}) &= p_1 \cdot p_2 \cdot p_3
\end{align}

For each OC, we performed the Markov chain Monte Carlo (MCMC) fits using the python package {\tt emcee} \citep{Foreman-Mackey2013}, using {\tt n\_walkers} $=32$,  {\tt n\_steps} $=10000$ and  {\tt burnin} $=5000$, and for $p$=6. With these conditions, the median and $16^{\rm th}$ and $84^{\rm th}$ percentiles of the $p+5$ likelihood parameters could be estimated for all 127 OCs from \citetalias{Cantat-Gaudin2020b}, and for 248 out of the 250 OCs from \citetalias{Tarricq2022}. 
In order to select only OCs for which the fitted polynomials and the Gaussian mixture model are accurate descriptions of the main-sequence CMD, we require the following (relatively strict) conditions:
\begin{enumerate}
    \item $\sigma_{\rm SS} \leq 0.2$,
    \item $\sigma_{\rm BS} \leq 0.25$,
    \item $|\Delta G + 0.75| \leq 0.05$,
    \item $\sigma_{f_b} \leq 0.25$ ($\sigma_{f_b}$ is the mean of ${f_b}$'s $16^{\rm th}$ and $84^{\rm th}$ percentiles),
    \item CMD is visually well fit (according to a visual inspection by the first three authors).
\end{enumerate}

The result of these cuts is a final sample of 202 OCs whose unresolved $f_{b}(q>q_{\rm lim})$ has been estimated using MCMC, 146 ($\sim 72\%$) from \citetalias{Tarricq2022} and 56 ($\sim 28\%$) from \citetalias{Cantat-Gaudin2020b}. 

Figure \ref{fig:mcmc_fit_corner_alessi1} shows an illustrative example for a cluster with a moderately populated MS, Alessi 1. We see that in this case (as in most other well-fit OCs; see Fig. \ref{fig:mcmc_examples}) apart from the polynomial coefficients there are little correlations between the fit parameters, in particular our main desirable parameter, $f_b$, is not strongly correlated with other parameters of the model, and the marginal posterior gives a sensible result ($f_b = 0.19^{+0.06}_{-0.05}$). We note that the significant correlations between the polynomial coefficients in the top left part of Fig. \ref{fig:mcmc_fit_corner_alessi1} are not a problem (provided that the single-star sequence is accurately fit by the polynomial), since we are primarily interested in the estimate of $f_b$.

Our method has three advantages: 1. the fit can be performed directly in observable space (the observed {\it Gaia} CMD), 2. it correctly incorporates uncertainties in both dimensions, and 3. the form of the fitted SS and BS functions is very flexible and does not depend on stellar models at all. In the vast majority of the cases, the adjusted polynomials give an accurate description of the observed single-star main sequences. The characterisation of the mass ratio threshold $q_{\rm lim}$ below which multiple systems are regarded as simple systems by our method is discussed in Sect. \ref{sec:MCMC_sims}.

\section{Application to simulated clusters}
\label{sec:sims}
 
In this section we describe the \emph{Gaia Object Generator} (GOG; \citealt{Luri2014}) and how we use it to generate realistic simulations of the OCs with well-determined $f_b$. We apply our MCMC method for deriving the unresolved $f_{b}(q>q_{\rm lim})$ to 10 realisations of the simulated CMDs of 324 OCs. We are thus able to estimate $q_{\rm lim}$ for the application of MCMC to both the simulated and observed CMDs. Then this $q_{\rm lim}$ value is used to estimate the resolved multiplicity fraction in the same $q_{\rm lim}<q \leq 1$ range as the unresolved one; and also the corresponding total multiplicity fraction $f_b^{\rm tot}(q>q_{\rm lim})$ (of both resolved and unresolved systems), for all the 202 OCs in our sample.

\subsection{Simulating clusters with the {\it Gaia Object Generator}}\label{sec:gog}

The \emph{Gaia Object Generator} (GOG; \citealt{Luri2014}) is a simulation tool that was developed to provide synthetic data that statistically reproduces the {\it Gaia} mission data
\citep{Gaia-Prusti}. For a given population of celestial objects, it applies the spacecraft and payload models to simulate the main {\it Gaia} observables (astrometry, photometry and spectroscopy) with realistic error models. 
Its complexity is increased after each {\it Gaia} data release, so that the simulated objects are in reasonable agreement with the data. 
GOG is typically used in conjunction with the {\it Gaia} Universe Model simulations (first described in \citealt{Robin2012}). 
The Universe Model relies on state-of-the-art descriptions of the characteristics of {\it Gaia} sources and on realistic scenarios for their formation, evolution, and dynamics. 
Most interestingly for our purpose, it includes a module that simulates multiple star systems \citep{Arenou2011}. 

For each OC we generate a synthetic population of single stars drawn from PARSEC 1.2S isochrones \citep{Bressan2012, Marigo2017} of solar metallicity with the age given by \citetalias{Cantat-Gaudin2020b}. The input stars are all single, generated according to the single-star IMF from \citet{Kroupa2001} and \citet{Kroupa2002}. Now, the  multiple-star module may change each input single star, according to some physically-motivated probability, into a system with the given star as primary and a lower-mass star of the same age as the secondary. Triples and higher order systems are not present in our simulations. The details of this process are described in \citet{Arenou2011}; in the next paragraphs we explain only the main assumptions.

The selection of single stars and primaries that will be part of a system is done so that they follow the luminosity function of primaries in the solar neighbourhood. For MS stars, the considered probability that an input single star gives birth to a system is given by the following function for the multiplicity fraction depending on the primary mass of the MS star, which is considered to fit well this dependence in the whole MS mass range: 
$$f_{b}(M_1)=83.88\cdot\tanh(0.688M_1 + 0.079).$$
This function is monotonically increasing with $M_1$, and is roughly compatible (by eye) with the several classes of dynamical decay models from \citet{Sterzik2004} or random pairing of \citet{Thies2007}. 
There is also abundant observational evidence for the increase of $f_{b}$ with $M_1$ \citep[e.g.][]{Kaczmarek2011, Sana2012, Fuhrmann2017}. 

\begin{figure}
\includegraphics[trim=0.51cm 0.8cm 1.8cm 1.4cm, clip,width=.49\textwidth]{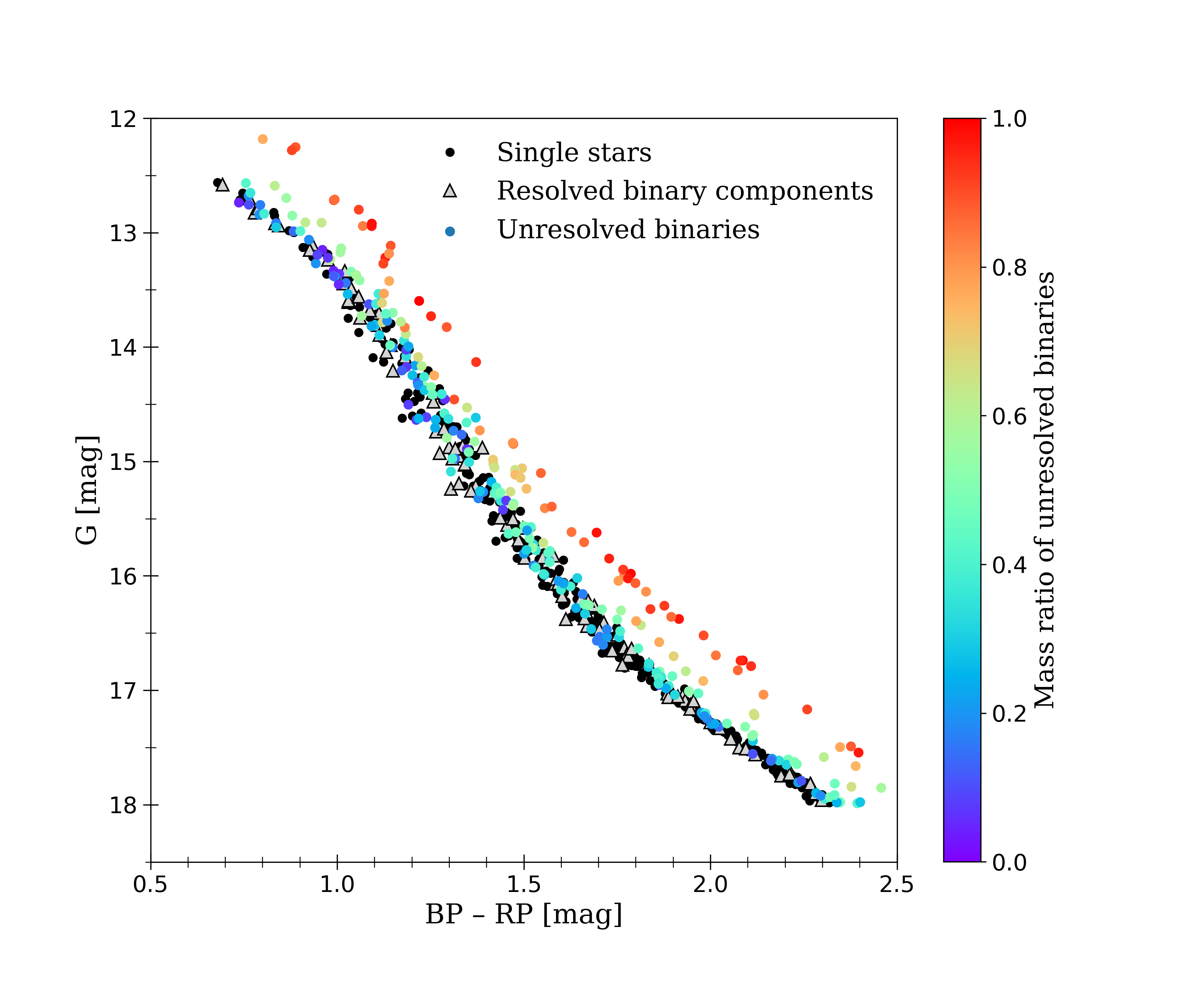}
\includegraphics[width=.415\textwidth]{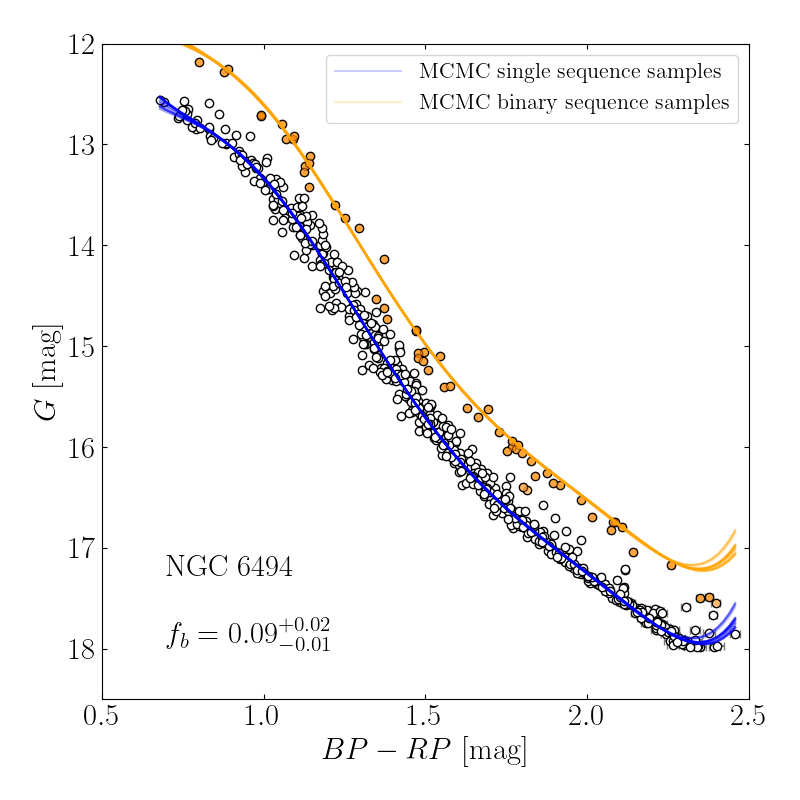}
\caption{One realisation of the simulated CMD of NGC 6494 (log age/yr $=8.58$, 682 selected main-sequence systems). 
\emph{Top panel:} Colour-coded by mass ratio. The black dots are single systems, unresolved binary systems are colour-coded by their mass ratio $q$. Light grey triangles with black contours overlapping with single stars are resolved components of binary systems. The errorbars are smaller than the markers. \emph{Bottom panel:} Result of the MCMC mixture-model fit, in the same style as Fig. \ref{fig:mcmc_examples}. The simulated $f_b$ for systems with $q>0.6$ for this realisation is 13\%. The mean unresolved binary fraction over the 10 CMD realisations is (14$\pm$7)\%.
}
\label{fig:cmd_sim_q_colorbar}
\end{figure}

For each generated multiple system, the mass of the secondary is drawn from the mass-ratio distribution $f(q)$, which is modelled as a probability density function that is linear by segment and depends on the spectral type of the primary and on the binary period. 
For the distribution of semi-major axis $a$, a Gaussian distribution in $\log(a)$ is assumed, with mean and standard deviation depending on $M_1$. From this random generation of $a$, the orbital period $P$ is drawn using Kepler’s third law. 
The eccentricities are assumed to be uniformly distributed within the interval $[0, 2E[e]]$, where $E[e]$ is the average eccentricity (which depends on $P$ and the primary’s spectral type). The other orbital parameters are then drawn randomly: the periastron date $T$ is chosen uniformly between $[0,P]$, the argument of the periastron ${\omega}_2$ uniformly in $[0, 2\pi]$, the position angle of the node $\Omega$ uniformly in $[0, 2\pi]$, and the inclination is chosen randomly in $\cos(i)$. A Roche model is used to avoid generating physically unrealistic systems of too small separations (for more details see \citealt{Arenou2011}).

Finally, GOG decides whether a multiple system can be resolved by {\it Gaia} (so that the components appear all along the same isochrone in distinct positions) or not (so that their flux is joined and the system appears above the single-star MS as a single point in the CMD). To do so, GOG takes into account the angular separation of the components of the system (we have adopted 500 mas, following \citealt{Fabricius2021}). Finally, GOG also adds magnitude-dependent photometric uncertainties.

\begin{figure*}[hbt!]
\vspace{4cm}
  \begin{overpic}[scale=0.45]{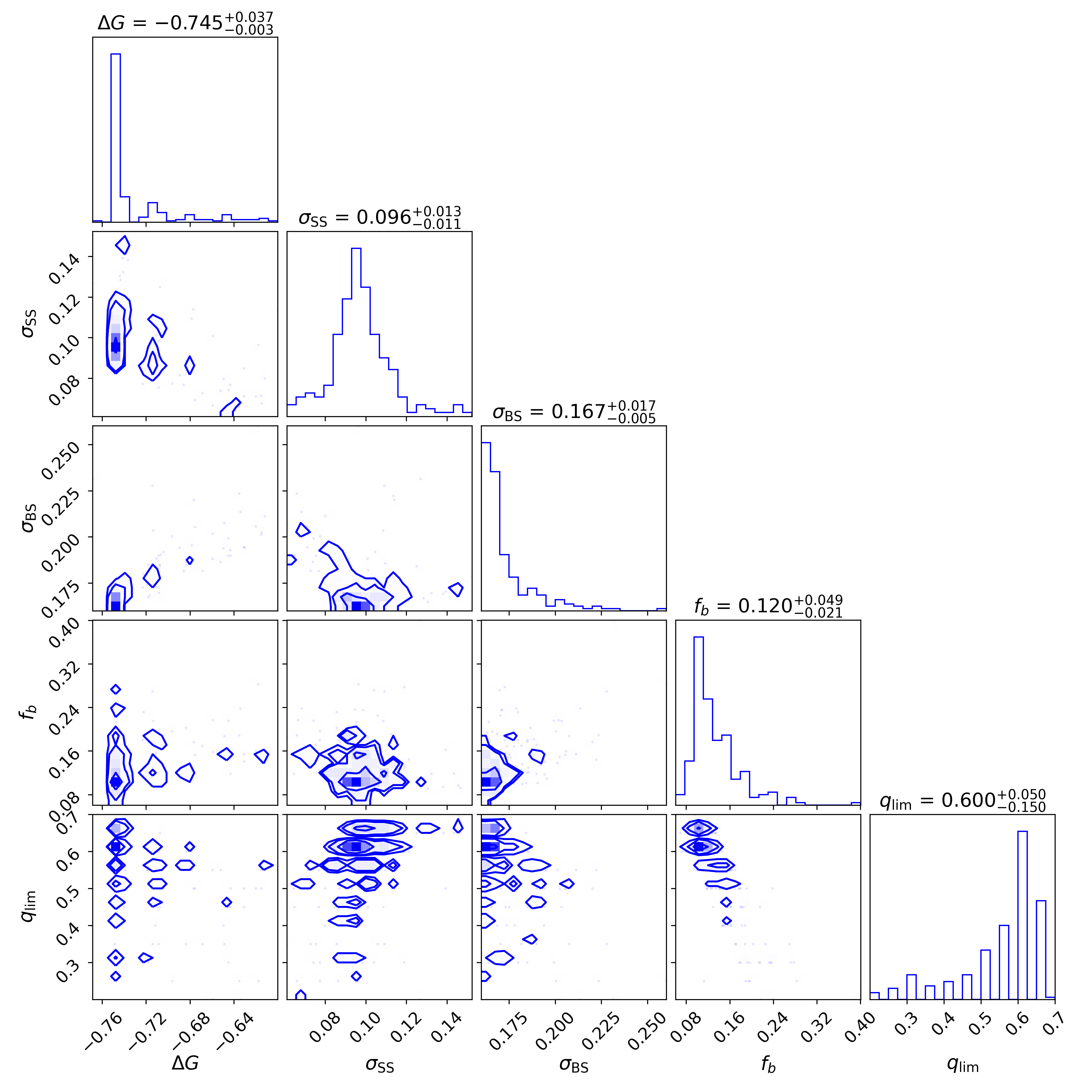}
     \put(62,50){\includegraphics[scale=0.41]{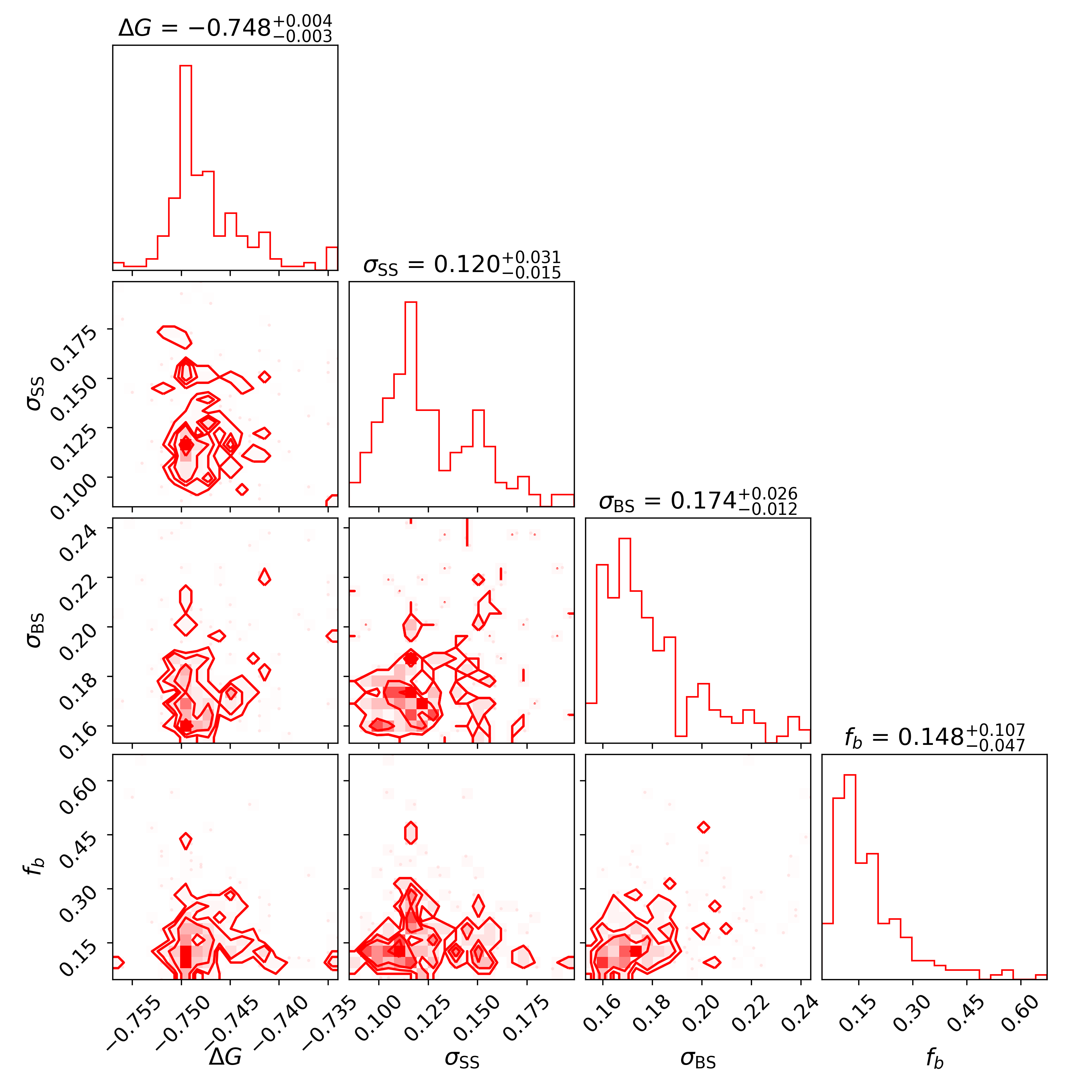}}  
  \end{overpic}
\caption{Summary {\tt corner} plots of the MCMC output parameters $\Delta G$, $\sigma_{\rm SS}$, $\sigma_{\rm BS}$, and $f_b$ for the cleaned results corresponding to the 219 GOG OC simulations (blue plots) and for the 202 well-fit {\it Gaia} OCs (red plots). In the case of the simulations, also $q_{\rm lim}$ is shown.}
\label{fig:cornerplots}
\end{figure*}

\subsection{Custom OC simulations}\label{sec:custom_sims}

For 324 OCs (corresponding to a preliminary sample of well-fit OCs) we have run custom simulations with the multiple-star module and GOG, applying them to each PARSEC single-star population of mass 10,000 $M_{\odot}$ with solar metallicity and the same age, distance, and extinction as the corresponding observed OC. The resulting synthetic population, now containing single stars and binaries, is returned by GOG as a catalogue of the members (or systems, if unresolved) with "true" and "observed" parameters. Photometric errors are taken into account, while other possible effects (most importantly differential extinction and rotation) that may contribute to the widening of the MS are not considered.

For each of these returned synthetic populations, we first select the MS members following the same procedure as for the observed OCs (see Sect. \ref{sec:ms_select}). 
To have the same number of members in the MS range as for the corresponding observed OC, we apply a random selection of MS members without replacement among all MS simulated members. 
A uniform probability in the random selection is adequate because the simulated population as seen by {\it Gaia} has already been populated according to an IMF for single stars, and then companions have been added. By doing so, the generated synthetic CMDs are approximate realisations of the corresponding OC. They automatically verify the conditions of having at least 30 MS members and a MS extension of 1 mag in colour index (see Fig. \ref{fig:cmd_sim_q_colorbar} for an example).
We generate $N_{sim}=10$ realisations of each observed OC using the same simulated data but applying $N_{\rm sim}$ different times the random selection of the observed number of selected MS members. This allows to take into account statistically the effect that the random selection of a reduced number of members can have in the inferred $f_b$.

\subsection{MCMC fitting to the simulated OCs}\label{sec:MCMC_sims}

We perform the equivalent estimations of $f_{b}$ through the mixture-model fitting described in Sect. \ref{sec:mcmc}, but now for a sample of 324 simulated OCs, 10 times each. We select those simulated OCs for which their CMDs have been well fit (according to the same criteria as in Sect. \ref{sec:mcmc}) for at least six out of the $N_{sim}=10$ realisations, and retain 219 (68\%) of the simulated sample  (144/219 are in common to the final sample of 202 observed OCs). The high-$q$ unresolved multiplicity fraction, $\overline{f_{b}^{\rm sim}}$, is calculated as the mean of the 10 values of $f_{b}^{\rm sim}(q>q_{\rm lim}^{\rm sim})$'s median obtained through MCMC; and its nominal uncertainty $\overline{\delta[{f_{b}^{\rm sim}}]}$ as the mean over the 10 realisations of the mean of its $16^{\rm th}$ and $84^{\rm th}$ percentiles. We also compute the standard deviation of $f_{b}^{\rm sim}$ over the 10 realisations (${\sigma}_{f_{b}^{\rm sim}}$); both values give similar results and decrease for OCs with more MS members as expected.

We can also derive additional information: as GOG first simulates the systems, and later decides whether they can be resolved by {\it Gaia} or not, we know which of our selected MS members are single stars, which are two resolved components of the same binary system, and which correspond to unresolved binary systems. Therefore, we can compute the mean of the theoretical unresolved $f_b$ over the 10 realisations: $\overline{f_{b}^{\rm sim, theo}}(q>0)$. We also compute the mean theoretical $f_b$ of unresolved multiple systems having a $q$ equal or greater than 11 different values: $\overline{f_{b}^{\rm sim, theo}}(q\geq q_{\rm min})$ (with $q_{\rm min}$ in the range [0.2, 0.7] in steps of 0.05). 
These values are then compared to $\overline{f_{b}^{\rm sim}}$, derived from the MCMC fits, to estimate $q_{\rm lim}^{\rm sim}$: $q_{\rm lim}^{\rm sim}$ is assumed to be equal to the $q_{\rm min}$ for which $\overline{f_{b}^{\rm sim, theo}}(q\geq q_{\rm min})$ is the closest possible to $\overline{f_{b}^{\rm sim}}$. 

For the sample with good MCMC fits, we determine a median value of $q_{\rm lim}^{\rm sim}=0.6_{-0.15}^{+0.05}$ (see Fig. \ref{fig:cornerplots}). Although $q_{\rm lim}$ is not the same for all clusters, we can confidently say that for a random observed and well-fit OC, we are determining the multiplicity fraction of systems with $q>0.6_{-0.15}^{+0.05}$.

\subsection{Estimation of the total multiplicity fraction: correction for resolved sytems}
\label{sec:resolved_correction}

One of the most nearby OCs is the Pleiades cluster. It was recently studied in detail by \citet{Torres2021} using \emph{Gaia} and long-term spectroscopic observations of thousands of stars. This work illustrates the complexity that one faces when trying to estimate the true $f_b$ of an OC, even in such a very nearby case. They find, after applying corrections for undetected binaries, a binary frequency (for periods up to 104 days) of $(25\pm3)\%$. When including known astrometric binaries, this estimate increases to more than 57\%.

In our study, we are limited to photometric observations and inherit the \emph{Gaia} magnitude limit from the membership catalogues. Our method provides an estimation of the unresolved $f_b$ of systems with $q>q_{\rm lim}$ based on their CMD positions, where $q_{\rm lim}$ is in principle characteristic of each OC. We therefore have to resort to simulations to correctly account for resolved systems, in order to be able to estimate the total $f_b$: $f_b^{\rm tot}=f_{b, \rm unres}+f_{b, \rm res}(d, age)$. As our unresolved $f_b(q>q_{\rm lim})=f_{b, \rm unres}^{\rm measured}$ is limited to systems of $q>q_{\rm lim}$, we also estimate $f_b^{\rm tot}$ of both unresolved and resolved binaries of $q>q_{\rm lim}$, which can be calculated as follows:
\begin{equation}
\label{eqn:total_BF_eq}
    f_b^{\rm tot}(q>q_{\rm lim}) = f_{b, \rm unres}^{\rm measured}·\left[1 + \frac {f_{b, \rm res}^{\rm sim}(q>q_{\rm lim}; \hspace{0.1cm} d, age)} {f_{b, \rm unres}^{\rm sim}(q>q_{\rm lim};\hspace{0.1cm} d, age)} \right]
\end{equation}

We estimate it using the ratio between the resolved and unresolved $f_b$ of the simulations instead of just adding up the measured unresolved $f_b$ with the simulated resolved $f_b$ in order to lessen the dependence on the total number of simulated binaries: what matters is the relative proportion between resolved and unresolved systems, not their absolute numbers.

\begin{figure}
\begin{center} 
\includegraphics[trim=0.5cm 0cm 0.6cm 0.2cm, clip,width=.49\textwidth]{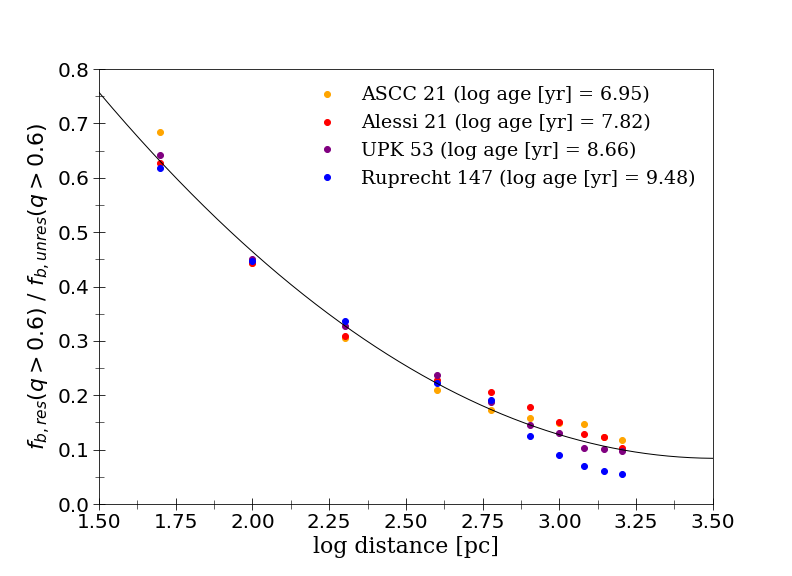}\\
\caption{
Ratio of the fraction of resolved binaries over the fraction of unresolved binaries, both of systems with $q>0.6$ $\left( \frac {f_{b, \rm res}^{\rm sim}(q>q_{\rm lim}=0.6;\hspace{0.1cm}  d)} {f_{b, \rm unres}^{\rm sim}(q>q_{\rm lim}=0.6;\hspace{0.1cm}  d)} \right)$, as a function of logarithmic distance, for four open clusters simulated with GOG. The black solid curve is the quadratic fit to the four open clusters that we use to compute the total multiplicity fraction.
} 
\label{fig:correct_dist}
\end{center}
\end{figure}

We use the GOG simulations to estimate the dependence of the resolved over unresolved multiplicity fractions as a function of the distance and age.  We simulate four OCs with ages covering the age range of the studied sample, placing them at 10 distances in the studied distance range, and compute their $f_{b, \rm res}^{\rm sim}$  over $f_{b, \rm unres}^{\rm sim}$ ratio for each distance, with a certain $q_{\rm lim}$ value. The result is represented in Fig. \ref{fig:correct_dist} for a value $q_{\rm lim}=0.6$ (justified and discussed in Sect. \ref{sec:MCMC_sims}). As expected, the closer the cluster, the higher the fraction of resolved binaries. There is no strong dependence on the OC’s age, so we fit a quadratic function to the dependence with the logarithm of the distance common for the four ages (black solid line in Fig. \ref{fig:correct_dist}). This function is used to interpolate the value of the ratio $\frac {f_{b, \rm res}^{\rm sim}(q>q_{\rm lim}; \hspace{0.1cm} d)} {f_{b, \rm unres}^{\rm sim}(q>q_{\rm lim};\hspace{0.1cm} d)}$ for each OC in our sample, and used to estimate its $f_b^{\rm tot}(q>q_{\rm lim})$.
We do not estimate the total $f_b$ of all binary systems because we would need to add the unresolved and resolved binaries with $q<q_{\rm lim}$, which requires a modelling of the $q$ distribution and depends on the specific properties of each binary system of the OC. 

\section{Results and discussion}\label{sec:results}

The results of our mixture-model fitting are included in a catalogue accessible via CDS. It contains the sample of 202 OCs well-fit by our mixture model (146 from \citetalias{Tarricq2022} and 56 from \citetalias{Cantat-Gaudin2020b}). We thus present one of the largest homogeneous catalogues of estimated multiplicity fractions in OCs to date.
For each OC, the catalogue provides the values and uncertainties of the unresolved multiplicity fraction deduced with MCMC and the estimated total multiplicity fraction (of both resolved and unresolved systems) of MS systems with $q>0.6$. The catalogue also contains the intrinsic colour range of the selected MS systems and its translation into masses; as well as the values of the other fit parameters. The most populated OC has 1158 MS members, the least populated one 30. Only 6 well-fit OCs have at least 500 MS members. The median mass of the least massive MS member selected is 0.71~M$_{\odot}$, and the median mass of the most massive MS member selected is 2.03~M$_{\odot}$.

The values of $f_{b}$ (before taking into account resolved binaries) that we find for the OC sample lie between $5\%$ and $67\%$, with a median value of 15\%. Their distribution (compatible with a log-normal distribution) is shown in the rightmost panel of Fig.~\ref{fig:cornerplots}. The typical uncertainties for $f_b$ amount to 4$\%$, and 95\% of the OCs have uncertainties smaller than 10\%.

A comparison with the simulations is  illustrated in Fig.~\ref{fig:cornerplots}. For the simulated OCs, we obtain values of  $\overline{f_{b}^{\rm sim}}\simeq(12^{+5}_{-2})\%$, with typical errorbars comparable to but often slightly lower than for the observed clusters. The simulations also allow us to determine the typical minimum mass ratio to which our method is sensitive: $q_{\rm lim}=0.6_{-0.15}^{+0.05}$. 

Our $q_{\rm lim}^{\rm sim}$ of $0.6$ is similar to the value found in the recent study of \citet{Jadhav2021}: for their considered magnitude range ($M_G \in [1, 10]$ mag; comparable to our range of selected MS members), they find $q=0.6$ to be the cut-off that ensures that the single MS stars are 3$\sigma$ away from the thus-defined BS.

We also estimate the total ($q>0.6$) multiplicity fraction, including resolved systems, for each OC (as explained in Sect. \ref{sec:resolved_correction}). For 89\% of all the 202 OCs, this correction of additionally taking into account resolved systems (apart from the unresolved binaries already accounted for in the unresolved multiplicity fraction) is smaller than 5\%. The total high-mass-ratio multiplicity fraction covers values from  6\% to 80\%, approximately following a log-normal distribution with a peak around 14\% and a median of 18\%. Only 9\% of the OCs in the sample have an $f_b^{\rm tot}(q>0.6) > 0.35$. The median total multiplicity fraction uncertainty is $\sim5\%$, and varies strongly from one OC to another.

\begin{figure*}[h!]
\centering
    \includegraphics[trim=0.5cm 0 0cm 0, clip, width=.45\textwidth]{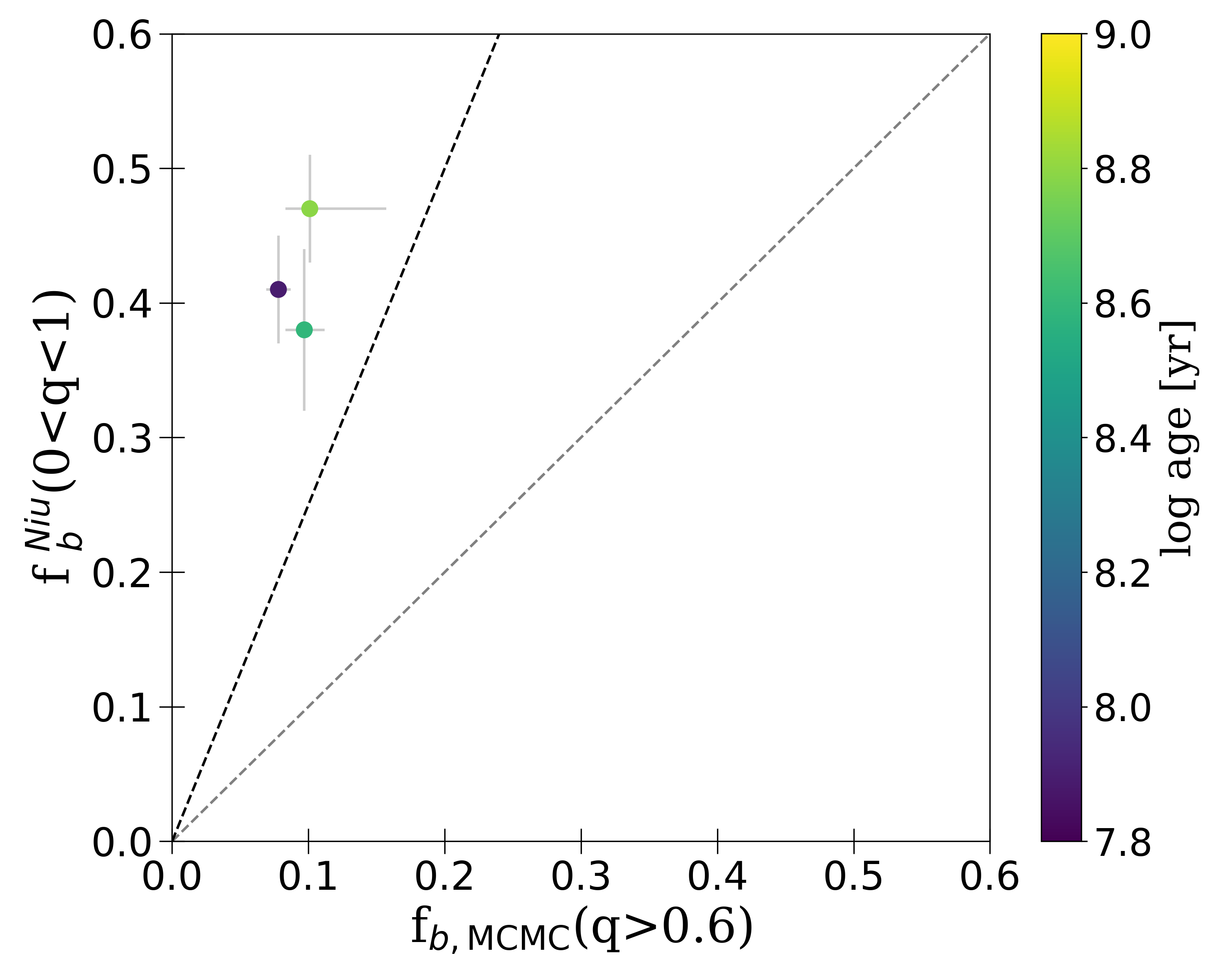}
    \includegraphics[trim=0.5cm 0 0cm 0, clip, width=.45\textwidth]{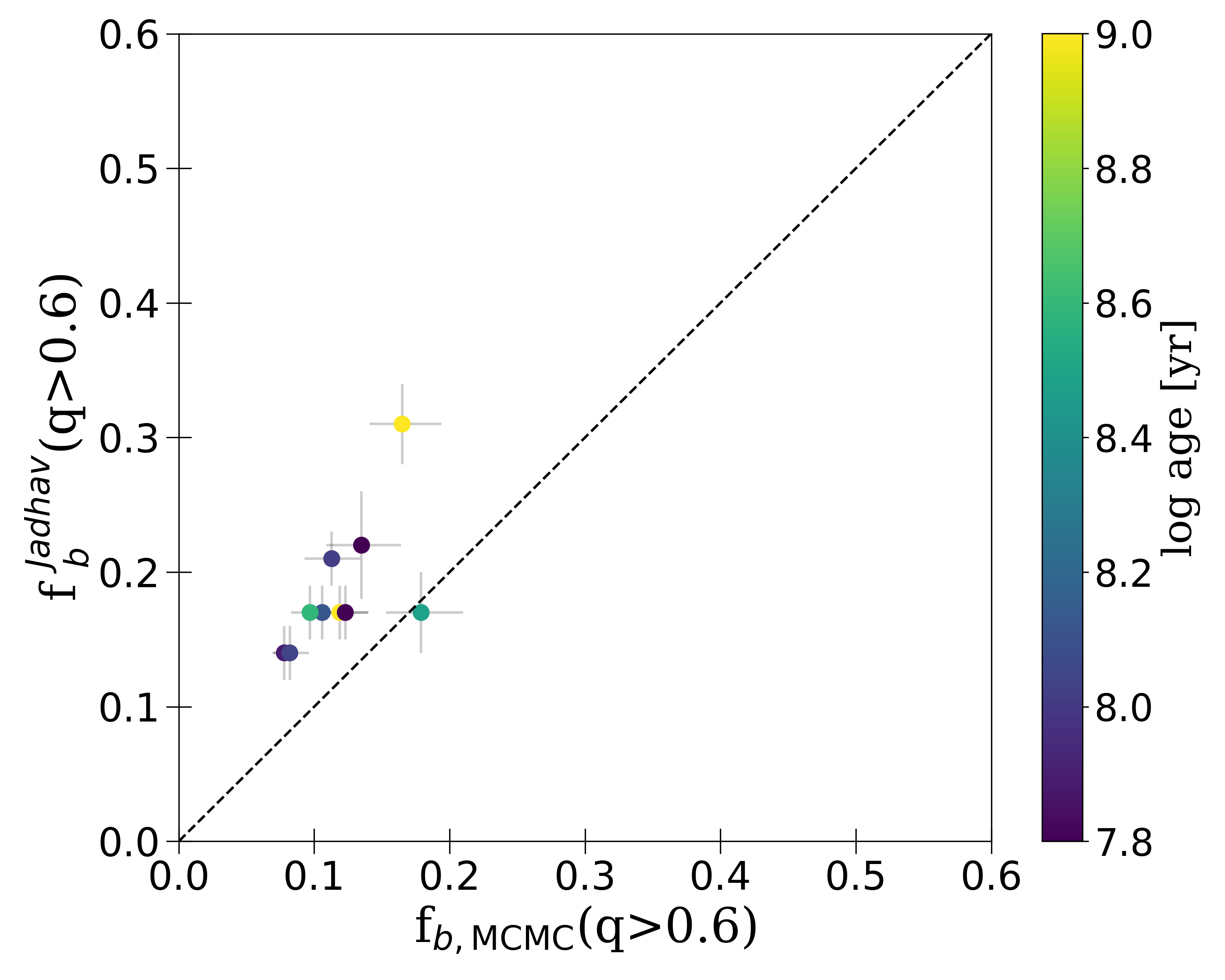}
    \includegraphics[trim=0.5cm 0 0cm 0, clip, width=.45\textwidth]{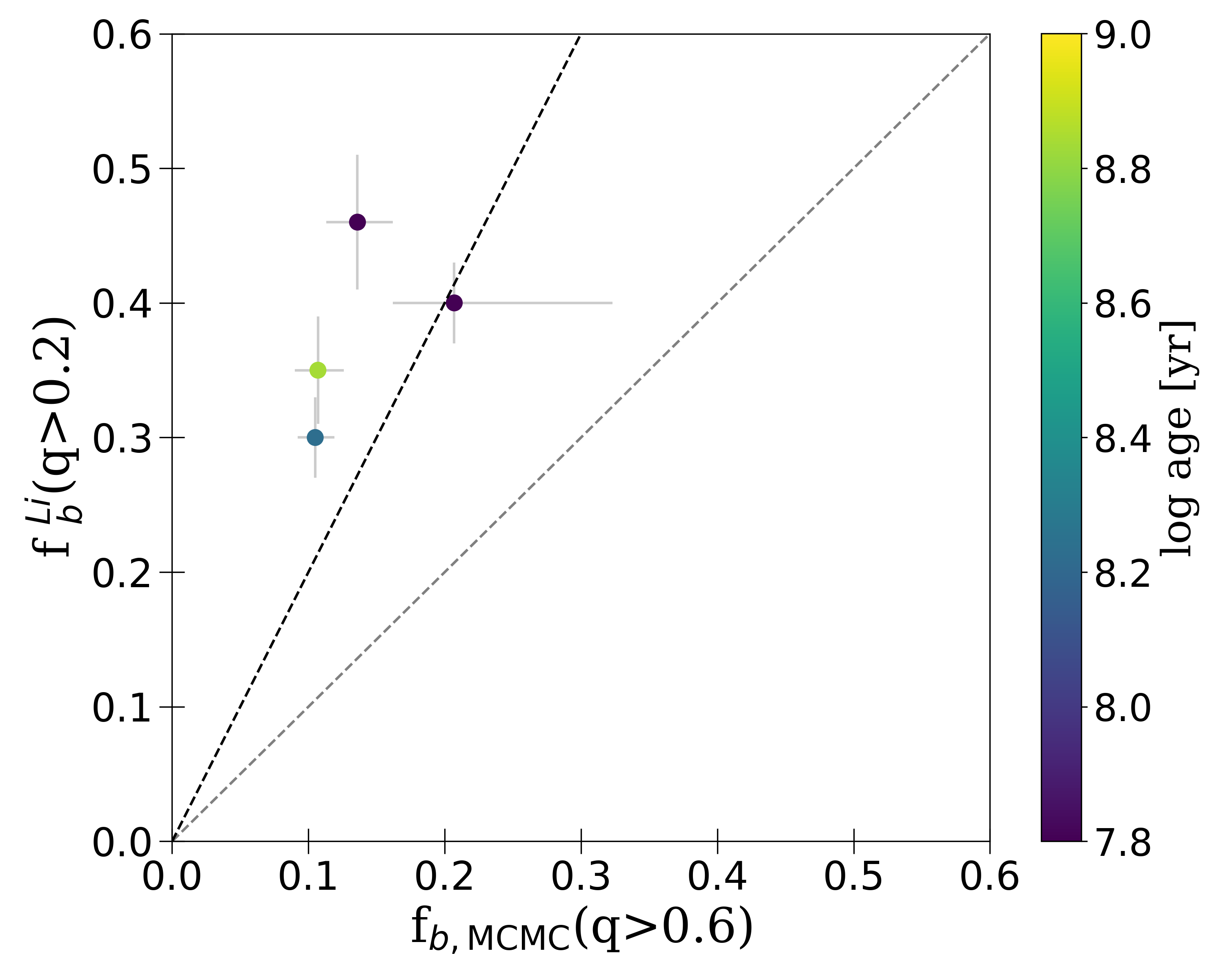}
    \includegraphics[trim=0.5cm 0 0cm 0, clip, width=.45\textwidth]{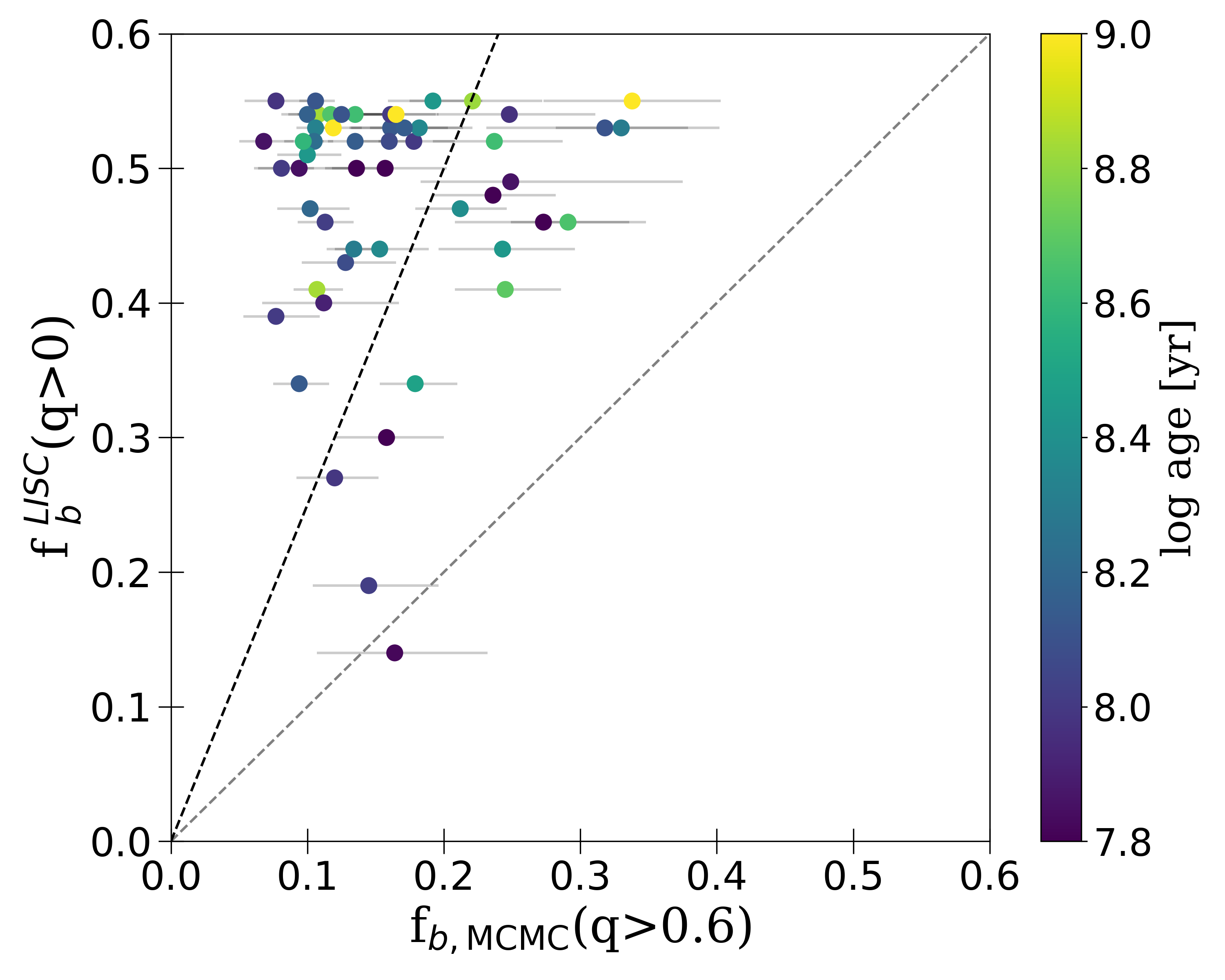}
\caption{Unresolved multiplicity fraction estimates from four literature studies as a function of our estimated unresolved multiplicity fraction $f_b(q>0.6)$, for the common open clusters between the two studies in each case. As a guidance for comparison, the equality line is drawn as a dashed grey line in each panel, and the black dashed lines show the expected relation when a flat mass-ratio distribution is assumed. {\it Top left:} \citet{Niu2020}. {\it Top right:} \citet{Jadhav2021}. {\it Bottom left:} \citet{LiShao2022}. {\it Bottom right:} \citet{Li2022}.}
\label{fig:comp_lit_4_figs}
\end{figure*}

\subsection{Comparison to other open cluster studies}\label{sec:literature}

The multiplicity fractions in many nearby OCs have been studied before \citep[e.g.][]{Sharma2008, Sollima2010, Cordoni2018, Cohen2020}; however, large homogeneous studies are still rare. The results of different studies, moreover, are very seldom directly comparable. Apart from the data and OC membership determinations used in each study, the main differences arise from the mass range covered by the studied OC members and the minimum $q$ to which the estimated $f_b$ is sensitive ($q_{\rm lim}$).

Besides, the estimated $f_b$ also depends on the assumptions of the modelling applied and the treatment of outliers. As discussed in \citet{DalTio2021}, the assumption that all binaries are unresolved is still a well-accepted approach to model binaries in CMD-fitting works. This approach generally provides $f_b$ values very similar to the ones found for more detailed prescriptions for the binaries for distant OCs. Since the fraction of resolved binaries decreases with distance (see Fig. \ref{fig:correct_dist}), the closer the studied OC, the less appropriate it is to blindly use the CMD-fitted $f_b$ assuming all binaries to be unresolved.

In this subsection we compare the results of our $f_b$ estimation with the literature for the OCs that we have in common with four of the most recent studies of $f_b$ in OCs. As they assume all binaries to be unresolved, we compare our likewise estimated $f_{b}$ with their results, instead of our calculated $f_{b}^{\rm tot}$ that takes into account the resolved binaries too.  \cite{Niu2020}, \cite{Jadhav2021} and \cite{LiShao2022} rely on the CMD of $G$ vs. $(BP-RP)$, as in our study, while \citet{Li2022} use $V$ vs. ($V$ - $I$). The comparisons are summarised in Fig. \ref{fig:comp_lit_4_figs}.

\citet{Niu2020} used {\it Gaia} DR2 photometry and LAMOST spectroscopy \citep{Zhao2012} to study the fundamental parameters of 12 well-populated OCs. From their fits they derived synthetic CMDs and inferred the $f_b$ of MS stars for systems with $0<q<1$. They found the $q$ distribution in their OCs to be flat, in accordance with the finding of \citet{Torres2021} for the Pleiades. They also provide the limiting values $m_{min}$ (minimum mass estimated by the faintest MS star in the OC) and $m_{max}$ (estimated by the brightest MS star in the OC) of the mass range for which $f_b$ has been estimated.
The top left panel of Fig. \ref{fig:comp_lit_4_figs} shows the comparison to our study for the three OCs we have in common. None of the studied mass ranges coincides and, as is shown in Fig. \ref{fig:comp_lit_4_figs}, our estimated $f_b$ are smaller than the ones obtained by \citet{Niu2020} in all cases. This is consistent with the fact that these authors both study mass ranges up to higher masses (while $m_{min}$ does not differ by much) and are sensitive to a much wider $q$ range, thus including more (low mass-ratio) binaries. To account for the latter effect, in each of the panels in Fig. \ref{fig:comp_lit_4_figs} the black dashed line shows the expected relation under the assumption of a flat mass-ratio distribution.

Furthermore, \citet{Niu2020} recalculate the $f_b$ for the Pleiades and NGC 2099 in mass ranges coincident with the ones of other published studies. We also perform our $f_{b}$ calculation in these two same mass ranges ($[0.6, 1.0]$ M$_{\odot}$ for the Pleiades and $[1.06, 1.63]$ M$_{\odot}$ for NGC 2099), so that we can compare our estimated $f_b$ for each OC with two different values computed in the same mass range but for different $q$ ranges. 

For the Pleiades (Melotte 22), our inferred $f_b(q>0.6)={0.086}^{+0.012}_{-0.011}$ is still smaller than the one of \citet{Niu2020}, $f_b(q<1)=0.20 \pm 0.03$, being compatible only within 3$\sigma$. This is mostly attributed to the fact that their minimum detected $q$ is lower, aside from intrinsic discrepancies. Our inferred $f_b$ is also much smaller than the one inferred by the pre-\Gaia study of \citet{Pinfield2003}, $f_b(0.5\leq q \leq 1)={0.23}^{+0.06}_{-0.05}$, despite being computed for a $q$ range compatible with ours.

For NGC 2099, our inferred $f_b(q>0.6)={0.11}^{+0.03}_{-0.03}$ is also smaller than the one of \citet{Niu2020}, $f_b(q<1)=0.23 \pm 0.02$, being compatible only within 3$\sigma$. Compared to the work of \citet{Cordoni2018}, however, our value and theirs ($f_b(0.7\leq q \leq 1)=0.085$) are computed for very similar $q$ ranges and are compatible within 1$\sigma$.

The top right panel of Fig. \ref{fig:comp_lit_4_figs} shows the comparison to \citet{Jadhav2021}, for the ten OCs we have in common. These authors used interpolated PARSEC isochrones to calculate the magnitudes of the primary star, secondary star and the combined unresolved binary for various $q$ values, and selected unresolved MS binaries of $q>0.6$ to compute the unresolved $f_b$ in 23 OCs using \Gaia DR2 data. They take into account, therefore, unresolved binaries in the same $q$ range as we do. They do not provide, however, the mass range for which $f_b$ is estimated, only the total number of MS members for each OC.

Their considered number of members is larger than ours for all OCs, although for some OCs they are comparable and for others our sample represents only a 30\% of theirs. For 9 OCs our estimated $f_b$ is smaller than theirs; while for only one OC (NGC 6793) ours is larger (although they are both compatible within 1$\sigma$).
These discrepancies can be real if the studied mass range is different for some of the OCs, or can otherwise arise (at least partially) from the fact that they refer to different OC members.

\citet{LiShao2022} apply a Bayesian framework that models the observed CMD of an OC as a mixture distribution of single stars, unresolved binaries and field stars, and then measure the fraction of unresolved binaries with $q>0.2$ among the member stars. They apply this method to 10 OCs with \Gaia EDR3 photometric data.
Figure \ref{fig:comp_lit_4_figs} (bottom left panel) shows the comparison to our study for the four OCs we have in common. Their $f_b$ is larger than ours for all OCs, which (as in the case of the comparison to \citealt{Niu2020}) is mainly due to the fact that they reach a lower $q$ threshold.

Finally, \citet{Li2022} determine the primordial $f_b$ fitting the CMD morphologies using the {\it Powerful CMD} code. They provide a catalogue, named LI team’s Star Cluster (LISC), for 309 OCs for which they consider the fit to be good, and for 288 OCs for which it is not as good, without specifying either the mass or $q$ ranges considered or the $f_b$ uncertainty. Our study and their good-fit one have 53 OCs in common. For all but one, their estimated $f_b$ is larger than ours and mostly incompatible with our values (see Fig. \ref{fig:comp_lit_4_figs}, bottom right panel). The study of \citet{Li2022} has the particularity of accumulating $f_b$ values between 0.5 and 0.55, which hints that their $f_b$ distribution might be mainly the result of an imposed prior and/or poorly constrained fits.

\begin{figure}
\includegraphics[width=.49\textwidth]{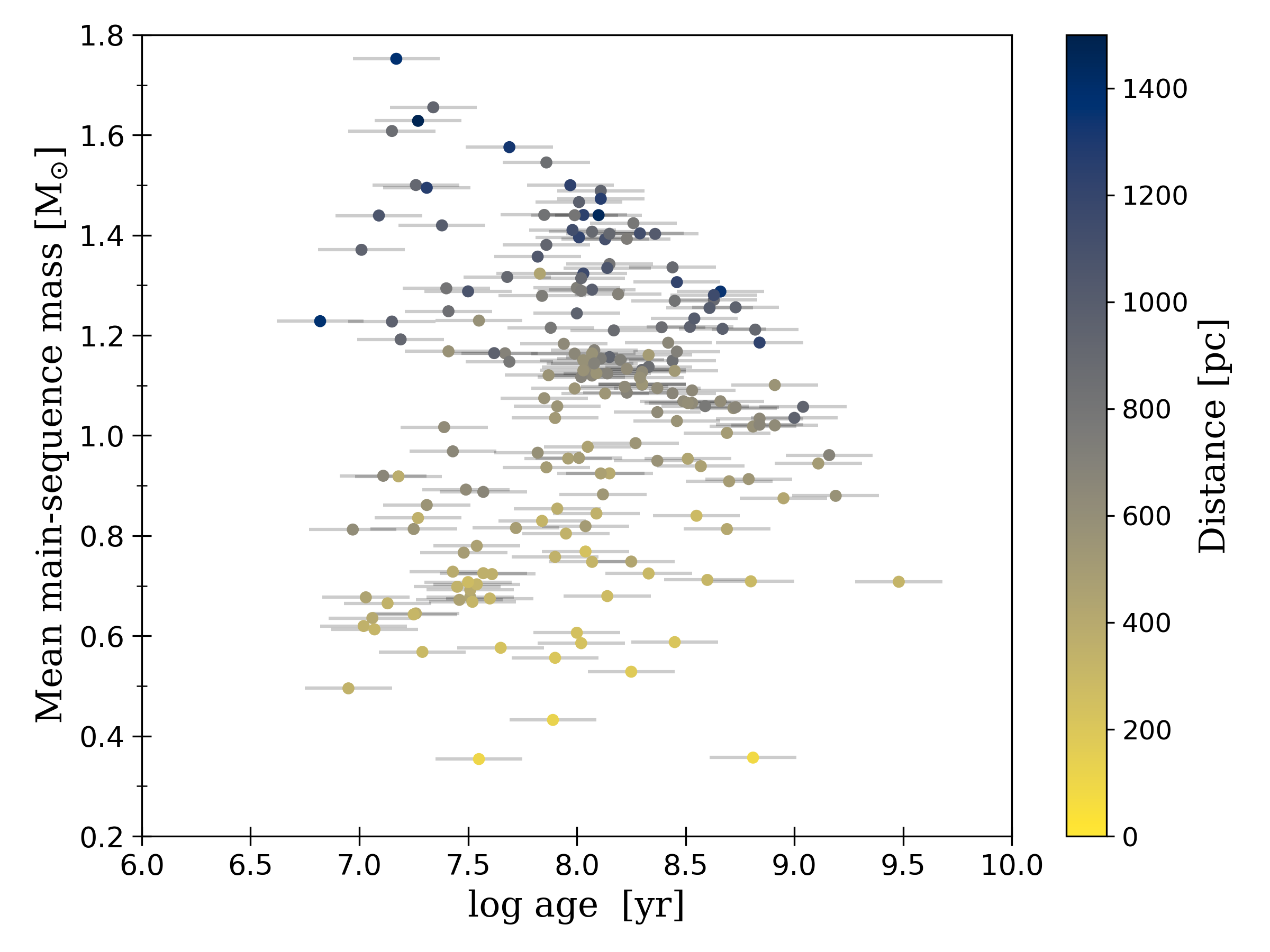}
\includegraphics[width=.49\textwidth]{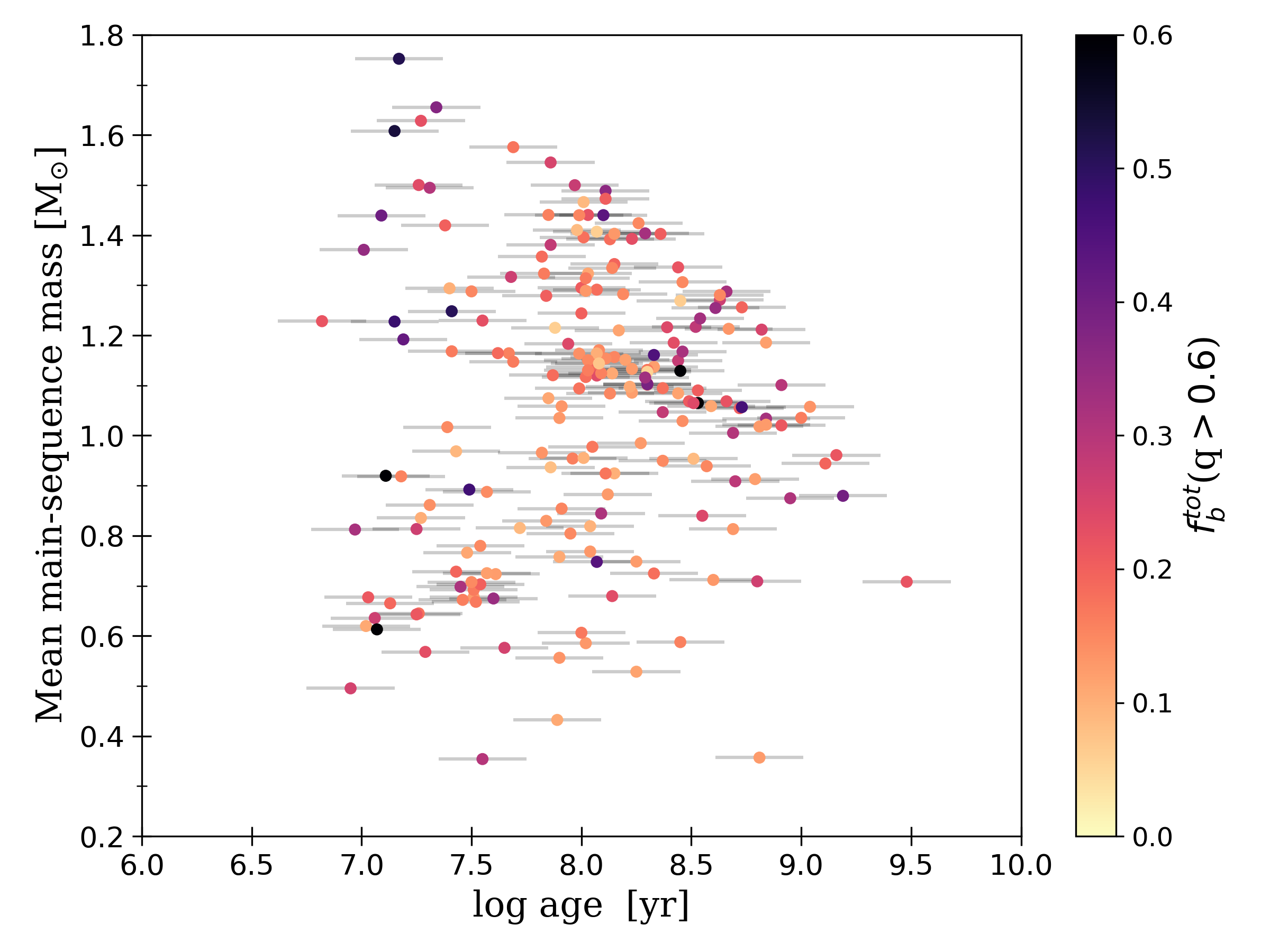}\\
\includegraphics[width=.49\textwidth]{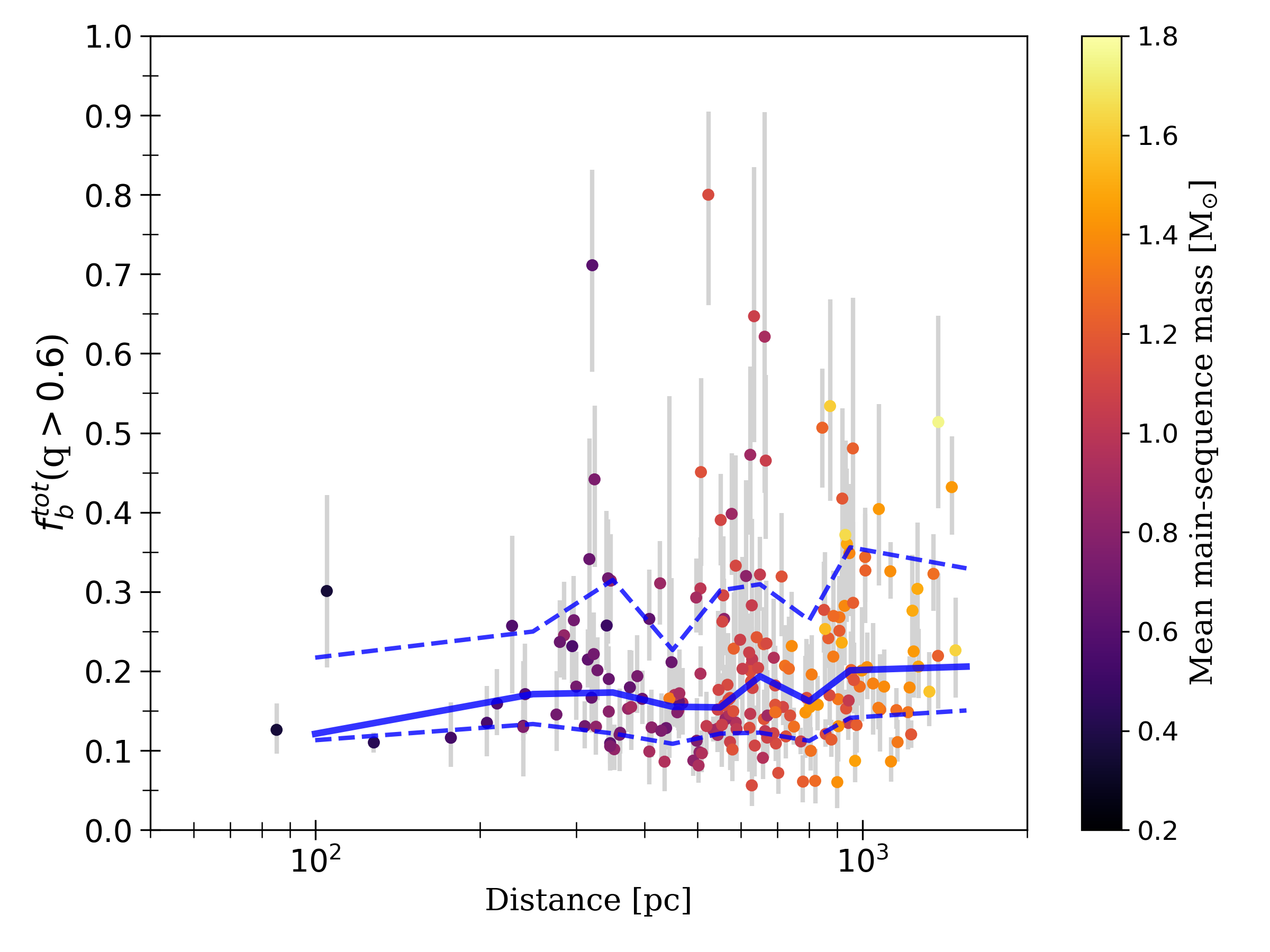}
\caption{Correlations between age, distance, mean main-sequence mass, and multiplicity fraction for the 202 studied open clusters. \emph{Top panels:} Mean main-sequence mass of the open cluster as a function of logarithmic age, colour-coded by their distance (top panel) and by their total multiplicity fraction of systems with $q>0.6$ (middle panel). \emph{Bottom panel}: Total multiplicity fraction of systems with $q>0.6$ as a function of logarithmic distance, colour-coded by the mean main-sequence mass of each open cluster. The central thick blue line is the running median, and the dashed blue lines are the $16^{\rm th}$ and $84^{\rm th}$ running percentiles.}
\label{fig:age_meanmass}
\end{figure}

\subsection{Dependence of the multiplicity fraction on the distance and stellar mass}\label{sec:biases}

Figure \ref{fig:age_meanmass} (top panel) demonstrates that for our distance-limited sample, the cluster parameters age, distance, and mean sampled mass are heavily entangled. We therefore expect that sample selection effects dominate most of the possible trends of the inferred multiplicity fraction with any cluster parameter (e.g. cluster mass, age, position in the Galaxy). The well-known dependence of the multiplicity fraction on stellar mass (see Sect. 2.2. of \citealt{Offner2022} and references therein), for example, is heavily imprinted in our sample, as we show below.

The bottom panel of Fig. \ref{fig:age_meanmass} shows the dependency of the estimated total multiplicity fraction with $q>0.6$ on distance. Firstly, we see that the number of studied OCs increases with distance as expected (due the larger volume covered), with only 4 OCs out of the 202 well-fit ones located closer than 200 pc.
Beyond this distance, there is a wide dispersion of $f_{b}^{\rm tot}$ values at any distance. Still, for both the unresolved and total multiplicity fractions, the running median displays a general tendency to increase with distance, because for more distant OCs we tend to see only the upper ends of their MSs, which have more massive stars and are thus expected to display a higher $f_{b}^{\rm tot}$. The increase of $f_{b}^{\rm tot}$ with the mass of the primary star is in full agreement with abundant observational evidence (see e.g. \citealt{Offner2022}, Figs. 1 \& 4). This effect is seen in the bottom panel of Fig. \ref{fig:age_meanmass}, where the colourbar corresponds to the mean MS mass of each OC, computed  as a function of the limits of our studied mass range (${m_{min}}$ and ${m_{max}}$) and \citet{Kroupa2001} power-law IMF ($\xi(m) \propto m^{{\alpha}}$ with ${\alpha}=2.3$ for $m \geq0.5$ ${\rm M}_{\odot}$):

\begin{equation}
\langle m \rangle = \frac{ \int_{m_{min}}^{m_{max}} m·\xi(m) \,dm}{ \int_{m_{min}}^{m_{max}} \xi(m) \,dm} = \frac{1-\alpha}{2-\alpha} \cdot \frac{{m_{max}}^{2-\alpha}-{m_{min}}^{2-\alpha}}{{m_{max}}^{1-\alpha}-{m_{min}}^{1-\alpha}}
\end{equation}

We find that for increasing distance, the observed ${m_{max}}$ does not vary significantly, while ${m_{min}}$ increases. For more distant OCs, only the upper MS ends are included in the magnitude-limited membership lists, thus preferring more luminous and massive stars (yielding thus a larger $\langle m \rangle$) and consequentially a higher $f_{b}^{\rm tot}$. Hence, the tendency of $f_{b}^{\rm tot}$ increasing with distance in the bottom panel of Fig. \ref{fig:age_meanmass} reflects the fact that the limiting apparent magnitude causes the observable portion of the OC’s MS to be dependent on the distance; and that the $f_{b}^{\rm tot}$ depends on the mass of the stars.

\begin{figure}[b!]
\begin{center} 
\includegraphics[trim=0.4cm 0.5cm 0.1cm 0.3cm, clip, width=.48\textwidth]{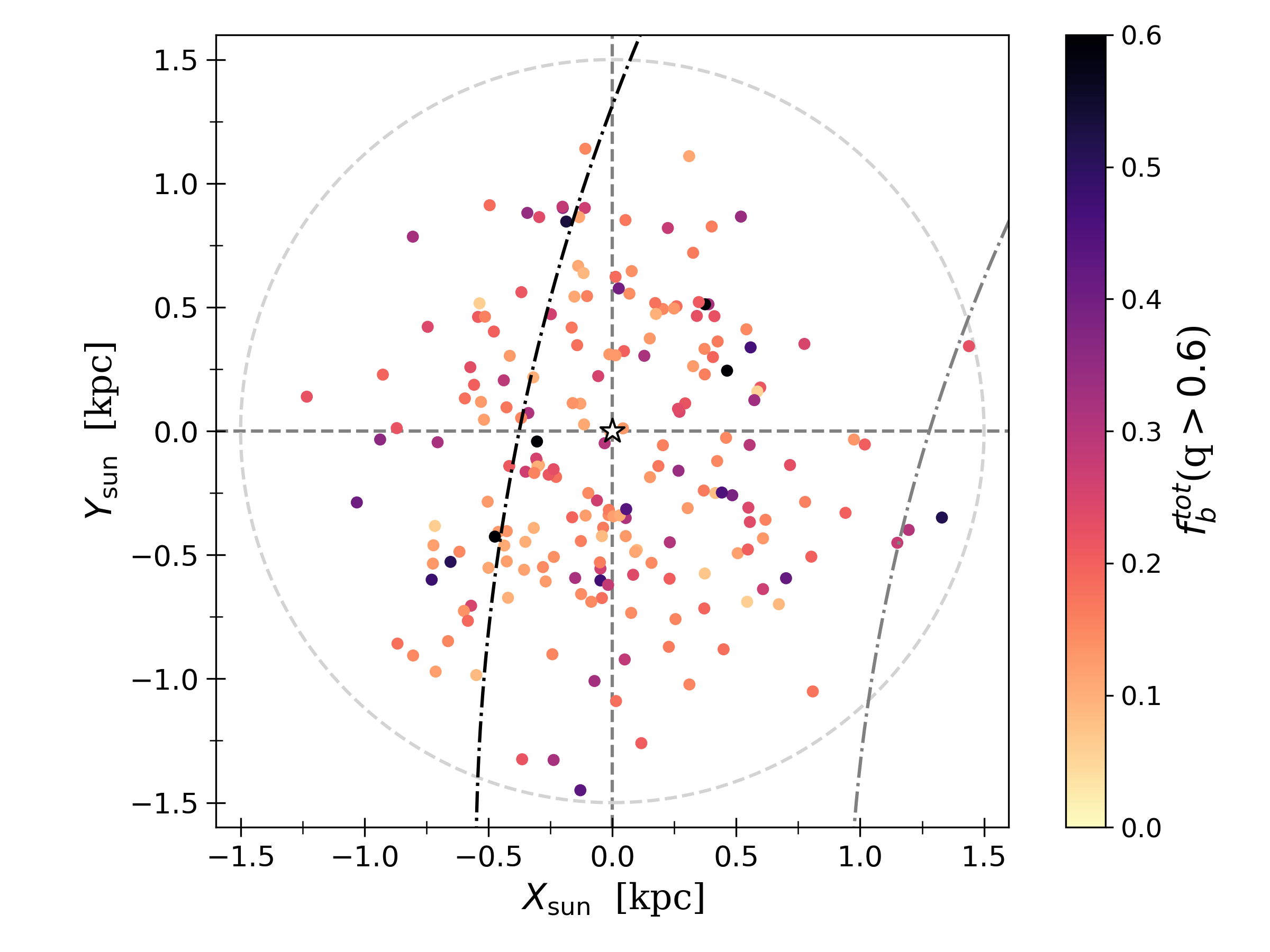}
\includegraphics[trim=0 0.5cm 0 0.3cm, clip, width=.48\textwidth]{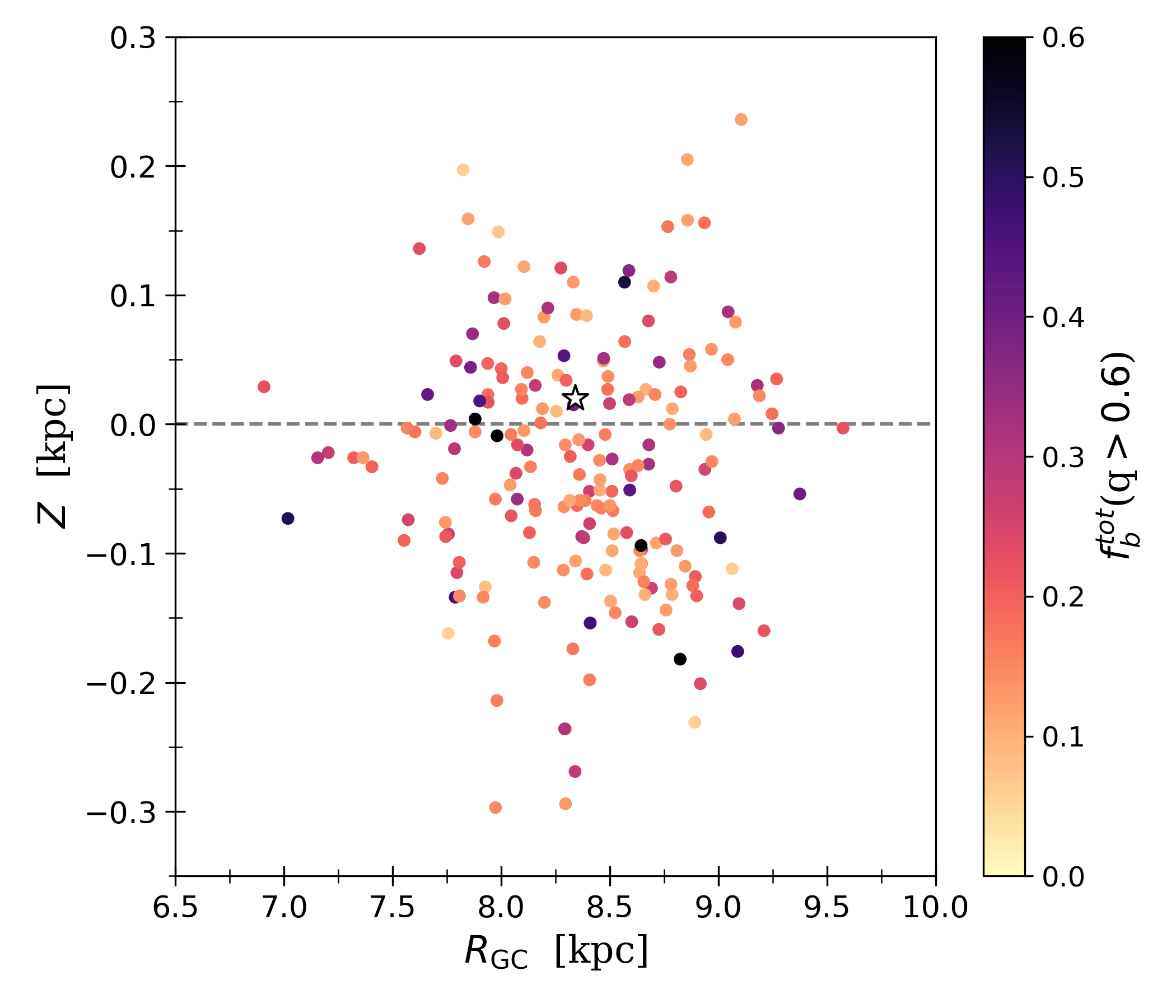}
\caption{
Locations of the 202 open clusters in our catalogue, colour-coded by the total high mass-ratio multiplicity fraction. \emph{Top panel}: top-down view on the Galactic plane (within a distance of 1.5 kpc, indicated by the dashed circumference). The Galactic centre is located towards positive $X_{\rm sun}$. The dash-dotted curves are the centres of the Local (black) and the Sagittarius–Carina arms (grey), as defined in \citet{Reid2019}. \emph{Bottom panel}: 
Height above the Galactic plane versus Galactocentric radius. The Sun’s position is indicated by an asterisk in both panels.
} 
\label{fig:fb_vs_position} 
\end{center}
\end{figure}

\subsection{Dependence of the multiplicity fraction on position}

\begin{figure*}[hbt!]
\begin{center} 
\includegraphics[trim=0 0 1cm 0.6cm, clip, width=.49\textwidth]{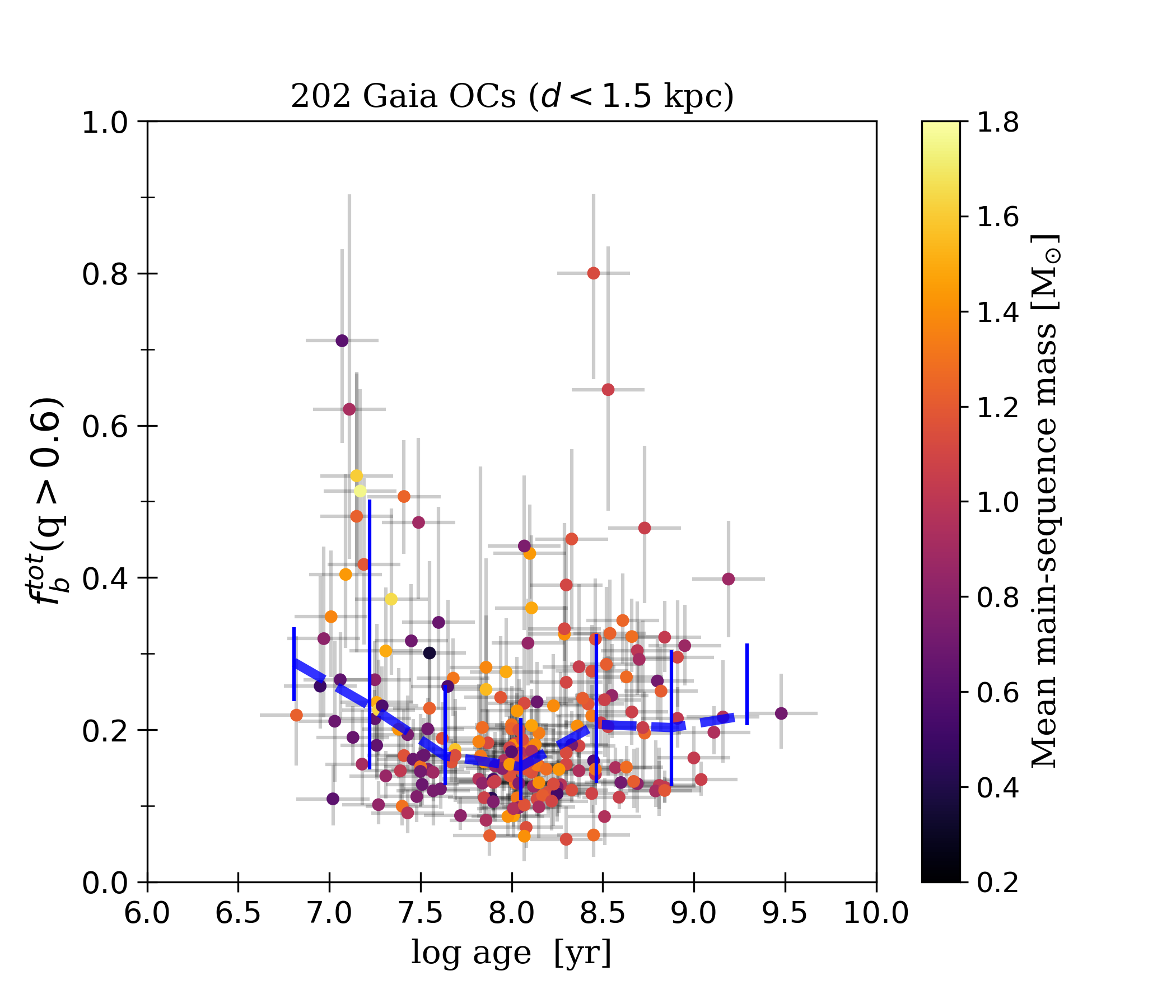}
\includegraphics[trim=0 0 1cm 0.6cm, clip, width=.49\textwidth]{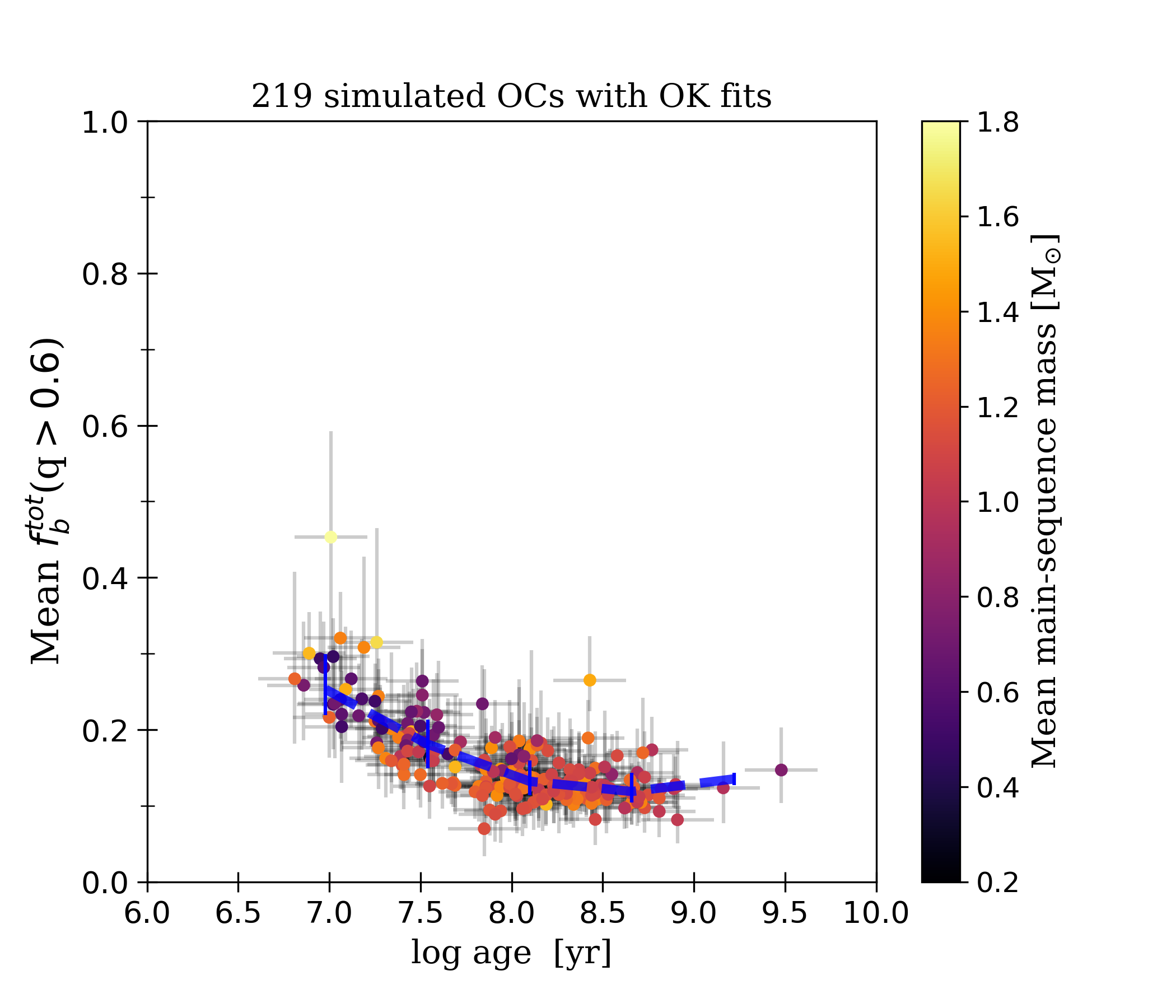}
\includegraphics[width=.49\textwidth]{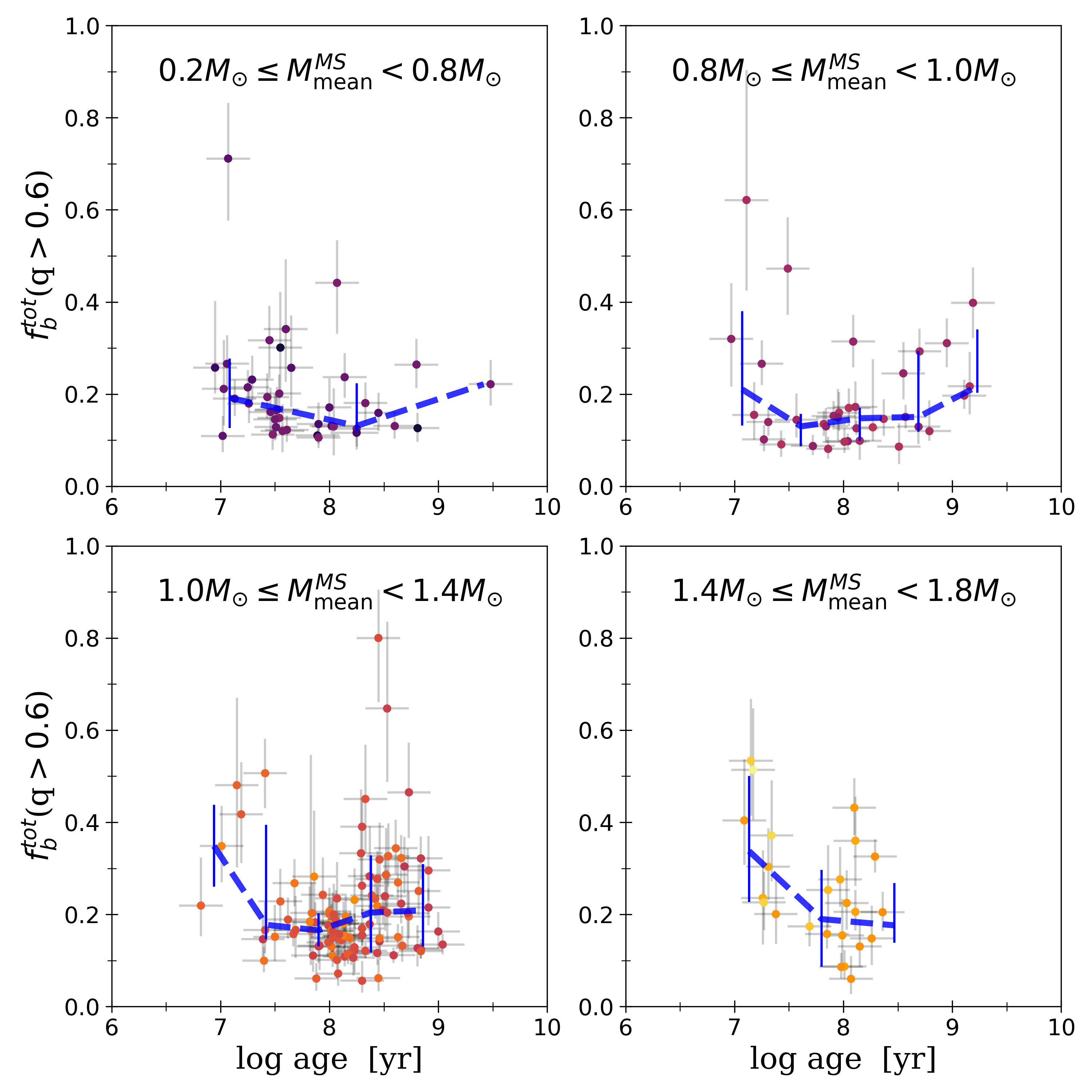}
\includegraphics[width=.49\textwidth]{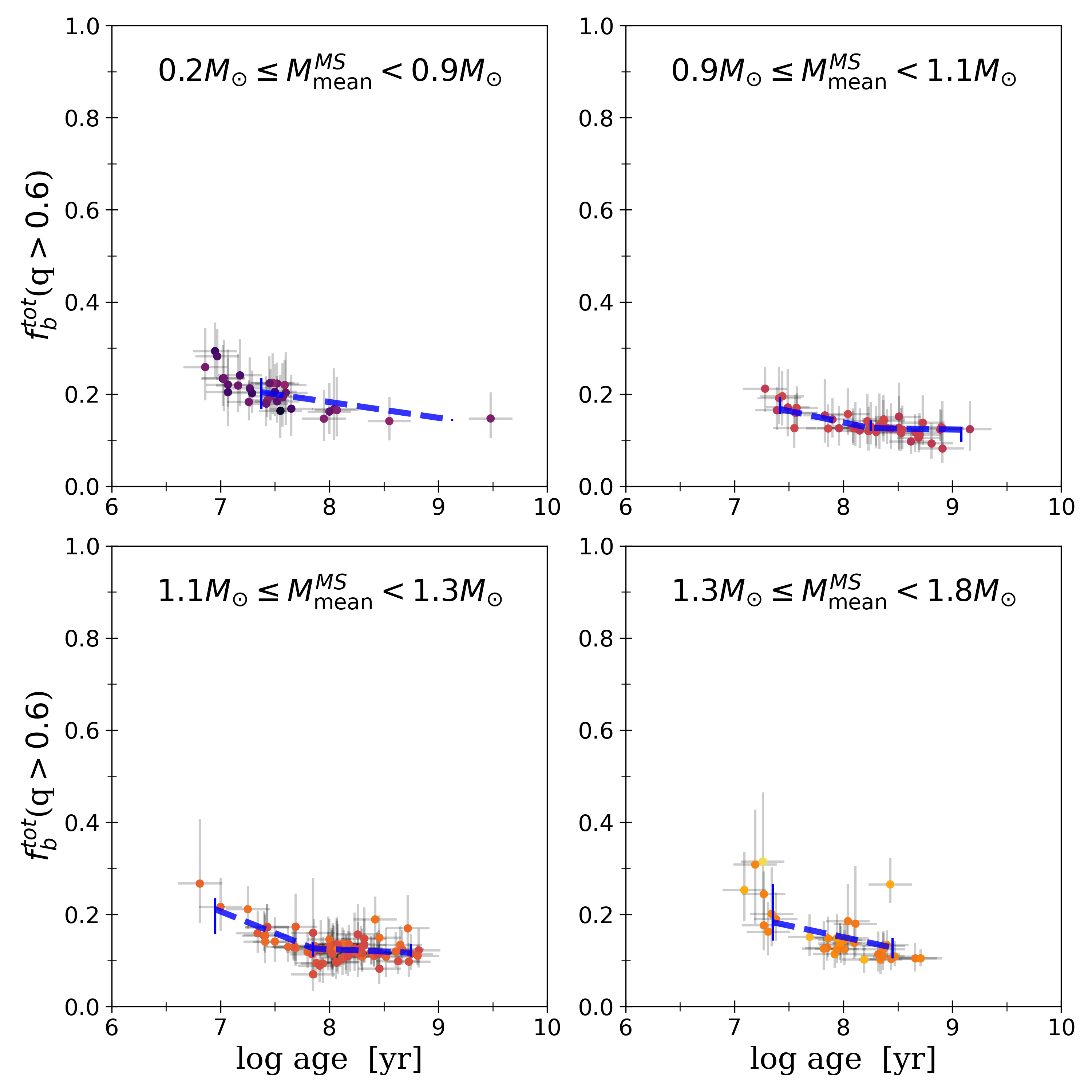}
\caption{
Total multiplicity fraction of systems with $q>0.6$ as a function of logarithmic age, colour-coded by the mean main-sequence mass of each open cluster. \emph{Top left panel}: For the 202 open clusters in our sample. The dashed blue line is the running median, and the errorbars are calculated as the $16^{\rm th}$ and $84^{\rm th}$ percentiles. \emph{Top right panel}: The same for the 219 well-fit simulated open clusters. \emph{Bottom panels}: The same as above, but now split into four bins of mean-sequence mass, both for the observed \Gaia OCs (left) and for the simulated clusters (right).
}
\label{fig:agedep} 
\end{center}
\end{figure*}

Figure \ref{fig:fb_vs_position} displays the three-dimensional spatial distribution of our OC sample colour-coded by $f_b^{\rm tot}(q>0.6)$. The top panel shows the projection onto the Galactic plane (similar to Fig. \ref{fig:xymap}). The OCs are approximately evenly scattered around the Sun, with a few of them being located close to the Local and the Sagittarius–Carina spiral arms. Apart from the distance dependence discussed above, no apparent correlation between the total multiplicity fraction and the Galactic position is seen. The bottom panel shows the OC’s height above the Galactic plane as a function of the distance from the Galactic centre. All the OCs fall inside the solar neighbourhood scale height for the thin disc, $z_d \approx$ 300 pc. As before, no clear correlation is seen between $f_b^{\rm tot}$ and the position in the Galactocentric frame.

\subsection{Dependence of the multiplicity fraction on age}\label{sec:age}

Figure \ref{fig:agedep} shows the dependence of $f_{b}^{\rm tot}(q>0.6)$ on age, both for the \Gaia sample and for its simulated counterpart. The distribution for the observed OCs (top left panel) is dominated by a large $f_b^{\rm tot}$ dispersion at all ages, which is not present in the simulated sample (top right panel) and therefore indicates a missing element of randomness in the GOG simulations. The running median decreases with age for the younger OCs up to approximately 100 Myr (log age/yr $\simeq8.0$). For OCs older than log age/yr $\simeq 8.0$ the total multiplicity fraction increases slightly with age. The age range for which we find an inversion of the trend is compatible with \citet{Thompson2021} who found a flattening of the unresolved multiplicity fraction trend with age at $\sim200$ Myr in a very small sample of 8 OCs.

The mean $f_{b}^{\rm tot}(q>0.6)$ vs. age distribution for the simulated OCs in the top right panel of Fig. \ref{fig:agedep} (where the mean $f_{b}^{\rm tot}(q>0.6)$ has been calculated with equation (\ref{eqn:total_BF_eq}) for $f_{b, \rm unres}^{\rm measured}=\overline{f_{b}^{\rm sim}}$) reproduces at least the decreasing trend for ages smaller than 100 Myr. Therefore, this trend is not a physical one, but mainly caused by the interplay of the mass dependence of the multiplicity fraction with the complex selection function of our OC sample. Each of the panels in Fig. \ref{fig:agedep} is colour-coded by the mean main-sequence mass of the portion of the MS considered by our fit. When dividing the sample into four bins of mean main-sequence mass $\langle m \rangle$ (lower panels of Fig. \ref{fig:agedep}), the age trends seen in the full sample persist. 

To better understand the trends seen in Fig. \ref{fig:agedep}, it is also helpful to return to the top panel of Fig. \ref{fig:age_meanmass}: for a certain age, the further the OC, the shorter its upper MS portion observed, which is thus more massive (increasing $\langle m \rangle$); whereas the closer the OC is, the observed MS extends from the upper end towards the less massive and more abundant stars that dominate (decreasing $\langle m \rangle$). Focusing on the further distances instead, the older the OC, the smaller $\langle m \rangle$. This reflects that the MSTO is shifted towards redder $BP-RP$ for older OCs, thus comprising less massive stars, and this is the reason why the top right corner of the top panel of Fig. \ref{fig:age_meanmass} is empty. 

From a dynamical-evolution point of view, a clear intrinsic dependence of the $f_{b}^{\rm tot}$ on OC age is also unexpected. For example, \citet{Hurley2007} showed with $N$-body simulations that the overall multiplicity fraction of a cluster "almost always remains close to the primordial value, except at late times when a cluster is near dissolution". 
Our simple GOG simulations (see Sect.~\ref{sec:sims}) also show that it is not necessary to invoke efficient multiplicity disruption during the early evolution of a cluster to explain the observed trend of $f_{b}^{\rm tot}$ decreasing with age during the first $\sim 100$ Myr.

\subsection{Dependence of the multiplicity fraction on the number of cluster members}\label{sec:number}

The physical quantities governing the dynamical evolution of a star cluster (modelled as a gravitational $N$-body system with stellar evolution) are the total mass, the initial mass function, the primordial multiplicity fraction, the central density, and the strength and variability of the external tidal field. Precise knowledge of most of these quantities is missing. However, it has been shown that clusters with similar remaining mass fractions (present-day mass divided by initial mass) have depleted their stellar mass functions by similar amounts of low-mass stars \citep{Baumgardt2003, Trenti2010}. \citet{Ebrahimi2022} have recently reinforced this result: they performed detailed synthetic CMD fitting (à la \citealt{Sollima2012}) for a sample of 15 nearby OCs and found a significant correlation between the present-day mass-function slope and the ratio of age to half-mass relaxation time.

\begin{figure}[h!]
\begin{center} 
\includegraphics[width=.49\textwidth]{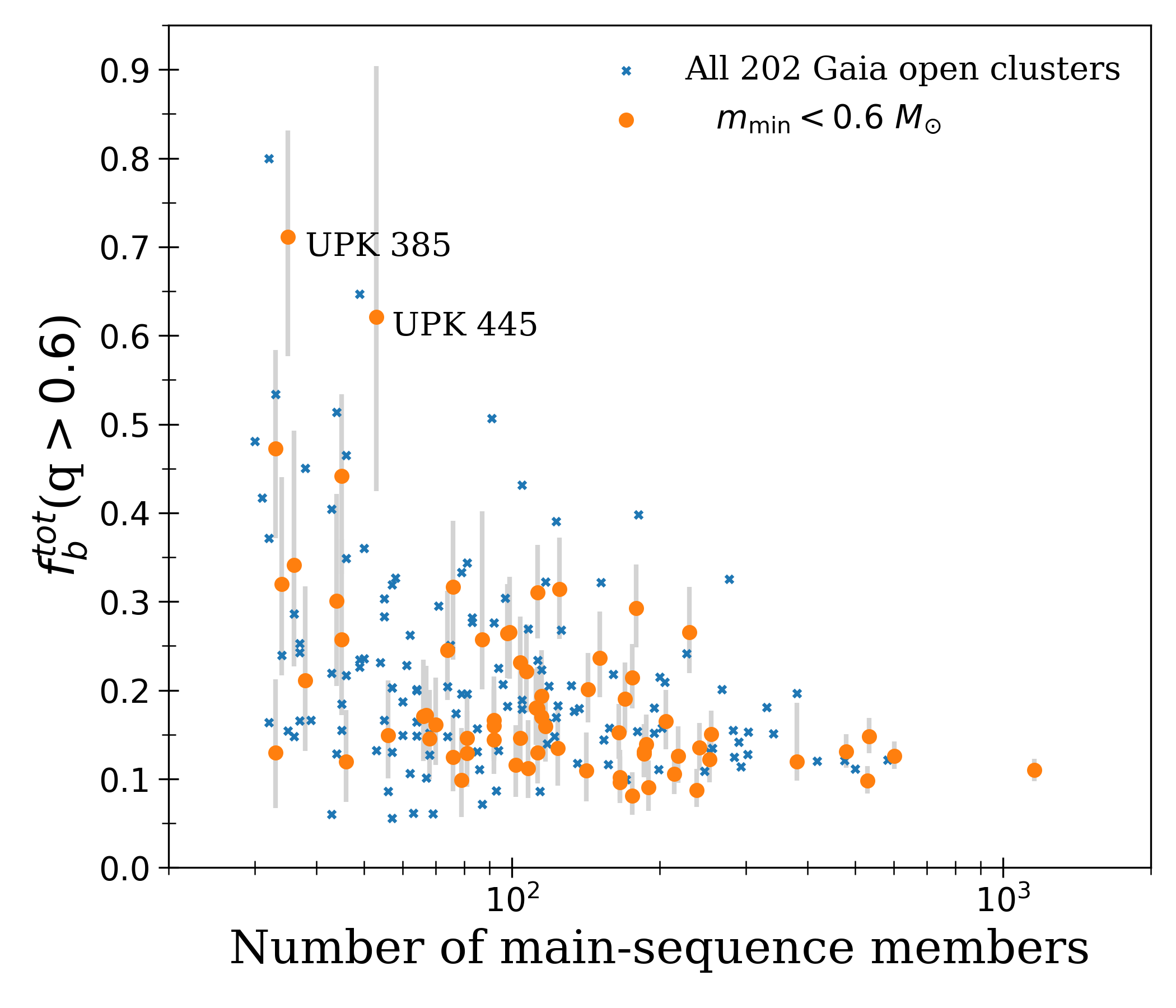}
\caption{Dependence of the total multiplicity fraction of systems with $q>0.6$ on the number of selected main-sequence members of each OC. OCs with extended low-mass main sequences ($m_{\rm min}<0.6$ M$_{\odot}$) are highlighted as big orange circles with errorbars. The names of the two OCs with extended low-mass main sequences and high $f_b^{\rm tot}$ (and thus possible candidates for close-to-disruption OCs) are annotated.}
\label{fig:nmemb} 
\end{center}
\end{figure}

In the absence of homogeneous estimates of the remaining mass fraction or the present-day mass function (our sample covers a too large distance range for this to be feasible), the number of MS members per OC can give a very rough estimate of how close an OC is to dissolution. Again, however, this number depends on the OC distance, age, and especially the completeness of the underlying membership catalogue, which can even vary from cluster to cluster \citep{Cantat-Gaudin2022}. In Fig. \ref{fig:nmemb}, we highlight OCs whose MSs extend to masses lower than 0.6 M$_{\odot}$ and show their total multiplicity fractions as a function of the number of main-sequence members. While it is difficult to detect a clear trend, we do find that the upper right part of the diagram is practically empty: for example, OCs with more than 200 members all have $f_b^{\rm tot}<0.5$ (or even 0.3 for the ones with $m_{\rm min}<0.6$ M$_{\odot}$), while for the least populated OCs we find cases with much larger total multiplicity fractions. The most interesting cases are the OCs with $m_{\rm min}<0.6$~M$_{\odot}$ and highly elevated multiplicity fractions, which could be candidates for being in their very final stage of evolution (UPK 385: $d=321$ pc, age $=12$ Myr, see Fig. \ref{fig:mcmc_examples}; UPK 445: $d=663$ pc, age $=13$ Myr; both from \citetalias{Cantat-Gaudin2020b}). A low initial velocity dispersion or a low initial central density in those clusters, or fewer interactions with the environment, could also lead to the observed high binary fractions \citep{Sollima2008}.

\subsection{Dependence of the multiplicity fraction on metallicity}\label{sec:metallicity}

Already the star-formation simulations by \citet{Machida2008} and \citet{Machida2009} suggested that low-metallicity gas clouds have a larger probability of fragmentation and thus a higher multiplicity frequency than metal-rich ones. Recent field-star analyses confirmed that the multiplicity fraction depends on the metallicity of the stellar system, at least for close binaries \citep{Badenes2018, El-Badry2019, Moe2019}. For example, \citet{Badenes2018} found, using APOGEE multi-epoch spectroscopy \citep{Majewski2017}, that metal-poor ([Fe/H] $\leq-0.5$) stars have multiplicity fractions $2-3$ times higher than metal-rich ([Fe/H] $\geq 0.0$) stars. The multiplicity fraction may further depend on the [$\alpha$/Fe] abundance ratio \citep{Mazzola2020}. 

Here we try to test these findings using the GSP-Spec metallicities \citep{Recio-Blanco2022} derived from the \Gaia DR3 data \citep{GaiaCollaboration2022Vallenari} for our parent sample, the OC members of \citetalias{Tarricq2022} and \citetalias{Cantat-Gaudin2020b}. We follow the recommendations of \citet{GaiaCollaboration2022Recio} and select a "high-quality" subset (see their Appendix B) with an additional effective temperature cut of $T_{\rm eff}<8000$ K. For 87 OCs in our sample GSP-Spec metallicities are available, but only for 23 of them we have metallicities for at least three member stars.

\begin{figure}
\begin{center} 
\includegraphics[width=.49\textwidth]{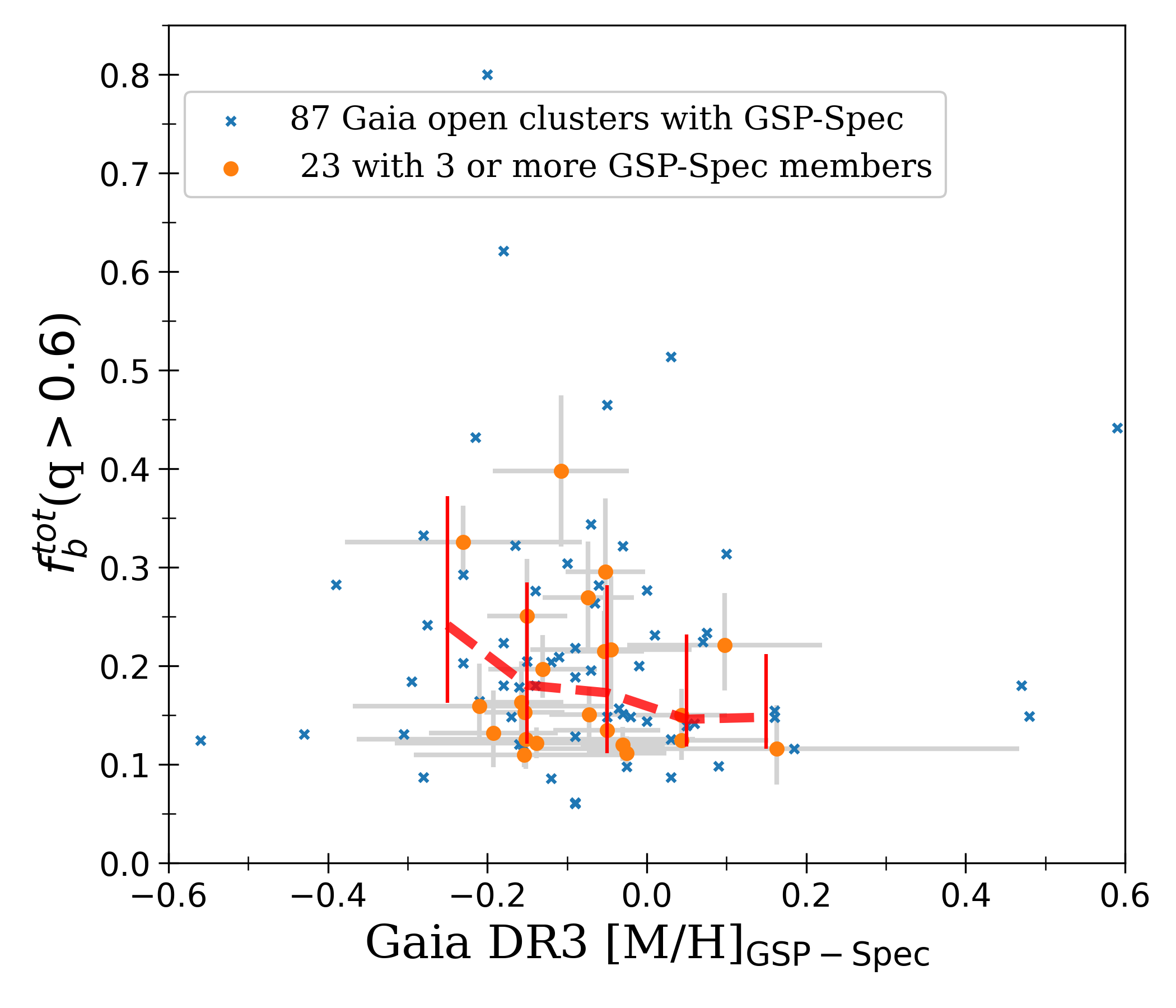}
\caption{Dependence of the open cluster's total multiplicity fraction of systems with $q>0.6$ on metallicity \citep[taken from][]{Recio-Blanco2022}. The 23 OCs with more than three individual GSP-Spec [M/H] measurements in \Gaia DR3 are highlighted as big orange circles with errorbars. The dashed red line is the running median for the 87 OCs in our sample for which GSP-Spec metallicities are available, and the errorbars are calculated as the $16^{\rm th}$ and $84^{\rm th}$ percentiles.}
\label{fig:metallicity} 
\end{center}
\end{figure}

Figure \ref{fig:metallicity} shows the result of this exercise. While we cannot clearly confirm that the total high-mass-ratio multiplicity fraction strongly depends on metallicity, due to the limited range in [M/H] sampled by our local OC sample, we see on average larger $f_b^{\rm tot}$ at sub-solar metallicities compared to super-solar metallicities, in agreement with the field-star literature \citep{Badenes2018, El-Badry2019, Moe2019}, confirming that our study is mostly sensitive to close binaries. In addition, our results indicate that the spread in $f_b^{\rm tot}$ is larger at [M/H]$<0$ than at [M/H]$>0$, although in the super-solar regime our number statistics is low. 

\section{Conclusions}\label{sec:conclusions}

The study of the multiplicity fraction $f_b$ in OCs requires membership determination and binary identification. Both have been made possible thanks to the unprecedented amount of high precision data for parallaxes, proper motions, and photometry provided by the successive data releases of the {\it Gaia} mission. In this paper we have successfully studied 202 {\it Gaia}-confirmed OCs closer than 1.5 kpc to the Sun, 146 from \citet{Tarricq2022} and 56 from \citet{Cantat-Gaudin2020b} catalogues, retaining only their members having a membership probability larger than $0.7$. Therefore, our conclusions are constrained by the membership completeness and selection biases in these works, as well as the incompleteness induced by the fit criteria of the present study.

The study of $f_b$ in clusters is usually restricted to the unresolved binaries. Their identification can be done individually through photometry (as done in this work and most of the above cited papers, e.g. \citealt{Niu2020, Borodina2021, LiShao2022}), but also through astrometry \citep{Belokurov2020, Penoyre2022, GaiaCollaboration2022Arenou} and radial-velocity monitoring \citep{Kobulnicky2014, Bodensteiner2021, Yalyalieva2022}.
Given the number of stars in each OC and the velocity precision necessary to detect most binary systems, spectroscopic surveys typically take many years to complete. Such individual studies are not feasible, therefore, to be carried out for large OC samples.

This is why the CMD, only requiring imaging in two filters, is still most often used for OCs as a fundamental diagnostic tool. A simple classification is usually done dividing the CMD into two regions, one associated with the SS locus (including single stars and low-$q$ systems) and the other with the equal-mass BS (including the higher-$q$ systems), as done in \citet{deGrijs2013}, \citet{Cordoni2018}, or \citet{Jadhav2021}. The degeneracy of metallicity and reddening, however, affects the isochrone fitting precision. An alternative is to create simulated CMDs using isochrones and binary prescriptions, and find the best fit (giving the nominal parameters) minimizing the distances between synthetic and observed data, as done in \citet{Sollima2010, Perren2015, Niu2020} or \citet{LiShao2022}.

Our approach, alternatively, models the observed $G$ vs. ($BP-RP$) CMD as a mixture distribution of single stars and unresolved binaries, considering that two Gaussian distributions centred on the SS and BS loci introduce the observed scatter in $G$ magnitude; one of them mostly accounting for the simple and low-$q$ systems, and the other for the high-$q$ binaries (and potentially higher-order systems). We are not concerned with estimating all the OC’s fundamental parameters, only the unresolved $f_b$, which is estimated as the weight of the BS's Gaussian component; without need for estimating the $q$ of each unresolved binary system. However, we do need to characterise the $q$ range the BS's Gaussian (and, thus, $f_b$) takes into account, which by construction involves the higher $q$ values down to $q_{\rm lim}=0.6_{-0.15}^{+0.05}$. We estimated this value by applying the method to a representative sample of simulated CMDs.

Applying strict criteria for a cluster to be well-fit, our MCMC implementation of the proposed mixture model yields good results for 202 (54\%) of the 377 nearby OCs with at least 30 MS members and a MS that extends over at least 1 mag in $BP-RP$, confirmed by visual inspection. It yields values of $f_{b}(q>0.6)$ in the range [0.05, 0.67] with a median nominal uncertainty of $0.04$ (not taking into account systematics). We also provide estimates of the total high-mass-ratio multiplicity fraction, $f_{b}^{\rm tot}(q>0.6)$, which takes into account the usually neglected portion of resolved binaries.

The main advantage of our modelling is that it does not rely on theoretical isochrone fitting. The versatility of the polynomial function fitted through MCMC enables us to study OCs with varied characteristics and degrees of differential extinction, thus enabling a homogeneous study of a large sample. The use of realistic {\it Gaia}-like simulations allows us to estimate both $q_{\rm lim}$ and the missing contribution from resolved binaries. 

The main drawbacks are: 1. All the OCs’ selected MS members are regarded as either simple or binary systems, not considering the possible presence of field star contamination or higher-order multiple systems; 2. The uncertainties intrinsic to the modelling are sometimes difficult to quantify. The uncertainty percentiles of the estimated $q_{\rm lim}=0.6_{-0.15}^{+0.05}$ could be considered too large for some purposes; 3. Our method only provides the multiplicity fraction integrated over the main-sequence mass range per OC, rendering a direct comparison to other studies difficult and resulting in complex selection effects (discussed in Sect. \ref{sec:biases}). For example, our unresolved $f_b$ estimations are coherent with the values obtained by \citet{Jadhav2021}, but much harder to be compared to other studies, which do not specify the considered mass and/or $q$ range (see Sect. \ref{sec:literature}). Overall, the $f_b$ estimation is by no means trivial, and it depends inevitably to some degree on modelling.

Summarising our main results from Sect. \ref{sec:results}, we find:
\begin{enumerate}
    \item The total high-mass-ratio multiplicity fraction taking into account also resolved binaries, $f_b^{\rm tot}(q>0.6)$, covers values from 6\% to 80\%, approximately following a log-normal distribution with a peak around 14\% and a median of 18\%. Only 9\% of the OCs in the sample have an $f_b^{\rm tot}(q>0.6) > 0.35$.
    \item $f_b^{\rm tot}$ increases with the mass of the primary star, in agreement with observational evidence from field stars and OCs \citep{Bouvier1997, Deacon2020}. 
    \item No apparent correlation appears between $f_b^{\rm tot}$ and the position in or perpendicular to the Galactic plane. 
    \item We observe a great dispersion for $f_b^{\rm tot}$ at all ages. There is, however, a decreasing trend with age until approximately $100$ Myr, followed by a slight increase for older clusters. 
    \item Our custom simulations using GOG show that part of the trend seen with age is caused by the complex selection effects (introduced by the mass dependence of the multiplicity fraction and the magnitude limit of our sample).
    \item The dispersion in $f_b^{\rm tot}$ seems to decrease significantly with the number of members (used as a proxy for cluster mass). We suggest that the highest values of $f_b^{\rm tot}$ (for UPK 385 and UPK 445) may indicate that those objects are very close to dissolution (in line with simulations by e.g. \citealt{Hurley2007}). Other possibilities such as a low initial velocity dispersion in those clusters, exist \citep{Sollima2008}. 
    \item The multiplicity fraction decreases with metallicity, in line with recent studies using close binaries in the field \citep{El-Badry2019, Moe2019}.
    \item Our results are available on the CDS. The code used to produce this paper is available on github\footnote{\url{https://github.com/fjaellet/oc_binary_fraction}}.
\end{enumerate}

The study of multiplicity is a vibrant field, especially in OCs and especially since the advent of {\it Gaia}. The recent review by \citet{Offner2022} nicely illustrates how multiplicity depends on a number of physical parameters, and that the determination of the overall multiplicity fraction is entangled with the determination of separation/period, eccentricity, and mass-ratio distributions, as well as with the chemical properties of stars. 
Future work is therefore necessary to corroborate our results with revised and more complete membership lists using the latest \Gaia data (including also radial velocities when possible; \citealt{GaiaCollaboration2022Arenou}), and to extend it to the determination of the mass-ratio distributions (ideally as a function of primary mass and not averaged over a whole OC). 

For nearby well-populated OCs, past works have already shown that it is possible to constrain the mass-ratio distribution (e.g. \citealt{LiShao2022}). The recent study of \citet{Albrow2022}, for example, shows that in M67 the mass-ratio distribution rises for $q > 0.3$, in line with an earlier study for NGC 188 \citep{Cohen2020}, but slightly at odds with the close-to-flat $q$ distribution found by \citet{Torres2021} for the Pleiades.

Another remaining question is whether the multiplicity fractions in OCs are measurably different from the multiplicity fractions in the field (except from the well-known effects of dynamical mass segregation; e.g. \citealt{Geller2012, Sheikhi2016, Tarricq2022, Casamiquela2022, Evans2022}). \citet{Parker2012} showed that stochastic interactions within an OC can significantly vary the intermediate-separation multiplicity fraction by a factor of 2. \citet{Parker2013} concluded, however, that the shape of the observed $q$ distribution is not significantly altered by internal interactions in (Orion-Nebula-type) clusters after its birth, but rather an imprint of the star formation process. How the process of multiple-star formation depends on the density (but also metallicity and magnetic fields) in the birth environment, remains an open question \citep{Offner2022}.

\makeatletter
\def\@fnsymbol#1{\ensuremath{\ifcase#1\or *\or \dagger\or \ddagger\or
   \dagger\or \mathparagraph\or \|\or **\or \dagger\dagger
   \or \ddagger\ddagger \else\@ctrerr\fi}}
\makeatother

\begin{acknowledgements}
\renewcommand{\thefootnote}{\fnsymbol{footnote}}
We thank Nigel Hambly (Edinburgh) who inspired the initial idea for this project by presenting a \Gaia DR2 {\tt python} tutorial at the Royal Astronomical Society's National Astronomy Meeting 2019 and Teresa Antoja (Barcelona) for suggesting to fit the main sequence and the multiplicity fraction simultaneously. 
We also thank Yoann Tarricq (Bordeaux) for making his OC membership catalogue available to us before publication. Finally, we thank the two referees, Antonio Sollima\footnote{{\url{https://www.media.inaf.it/2023/02/05/ricordando-antonio-sollima/}}} and an anonymous referee, for constructive comments that helped improve the paper. 

This work has made use of data from the European Space Agency (ESA) mission \textit{Gaia} (www.cosmos.esa.int/gaia), processed by the \textit{Gaia} Data Processing and Analysis Consortium (DPAC, www.cosmos.esa.int/web/gaia/dpac/consortium). Funding for the DPAC has been provided by national institutions, in particular the institutions participating in the \textit{Gaia} Multilateral Agreement. 

This work was partially funded by the Spanish MICIN/AEI/10.13039/501100011033 and by "ERDF A way of making Europe" by the “European Union” through grant RTI2018-095076-B-C21 and PID2021-122842OB-C21, and the Institute of Cosmos Sciences University of Barcelona (ICCUB, Unidad de Excelencia ’Mar\'{\i}a de Maeztu’) through grant CEX2019-000918-M. 
FA acknowledges financial support from MICIN (Spain) through the Juan de la Cierva-Incorporación programme under contracts IJC2019-04862-I and RYC2021-031638-I (the latter co-funded by European Union NextGenerationEU/PRTR). MG acknowledges support from grants EUR2020-112157, PID2021-125485NB-C22 and SGR-2021-01069 (AGAUR).

The preparation of this work has made use of TOPCAT \citep{Taylor2005}, NASA's Astrophysics Data System Bibliographic Services, as well as the open-source Python packages \texttt{astropy} \citep{Astropy2018}, \texttt{NumPy} \citep{VanderWalt2011}, \texttt{corner} \citep{Foreman-Mackey2016}, 
and \texttt{emcee} \citep{Foreman-Mackey2013}. The figures in this paper were produced with \texttt{matplotlib} \citep{Hunter2007}. 

\end{acknowledgements}

\bibliographystyle{aa} 
\bibliography{oc_binaryfraction}

\end{document}